\documentclass[journal]{IEEEtran}
\usepackage[ruled,linesnumbered]{algorithm2e}
\usepackage{cite}
\usepackage{amsmath,amssymb,amsfonts}
\usepackage{algorithmic}
\usepackage{graphicx}
\newtheorem{proposition}{\underline{Proposition}}
\newtheorem{remark}{\underline{Remark}}
\newtheorem{lemma}{\underline{Lemma}}
\usepackage{textcomp}
\usepackage{xcolor}
\usepackage[hidelinks]{hyperref}
\usepackage[utf8]{inputenc}
\usepackage{color}
\def\BibTeX{{\rm B\kern-.05em{\sc i\kern-.025em b}\kern-.08em
T\kern-.1667em\lower.7ex\hbox{E}\kern-.125emX}}
\begin{document}
\newcommand{\mv}[1]{\mbox{\boldmath{$ #1 $}}}

\title{Optimal Beamforming for Secure Integrated Sensing and Communication\\ Exploiting Target Location Distribution\\
\thanks{Part of this paper was presented at the IEEE Global Communications Conference (Globecom),  Kuala Lumpur, Malaysia, Dec. 2023\cite{bibGlobecom}.}
\thanks{The authors are with the Department of Electrical and Electronic Engineering, The Hong Kong Polytechnic University, Hong Kong SAR (e-mails: kaiyue.hou@connect.polyu.hk; shuowen.zhang@polyu.edu.hk).}
}
\author{\IEEEauthorblockN{Kaiyue Hou and Shuowen Zhang}}
\maketitle

\begin{abstract}
In this paper, we study a secure integrated sensing and communication (ISAC) system where one multi-antenna base station (BS) simultaneously communicates with one single-antenna user and senses the location parameter of a target which serves as a potential eavesdropper via its reflected echo signals. In particular, we consider a challenging scenario where the target's location is \emph{unknown} and \emph{random}, while its distribution information is known \emph{a priori} based on empirical data or target movement pattern. First, we derive the \emph{posterior Cram\'er-Rao bound (PCRB)} of the mean-squared error (MSE) in target location sensing, which has a complicated expression. To draw more insights, we derive a tight approximation of the PCRB in \emph{closed form}, which indicates that the transmit beamforming should achieve a ``\emph{probability-dependent power focusing}'' effect over possible target locations, with more power focused on highly-probable locations. Next, considering an artificial noise (AN) based beamforming structure at the BS to alleviate information eavesdropping and enhance the target's reflected signal power for sensing, we formulate the transmit beamforming optimization problem to maximize the worst-case secrecy rate among all possible target (eavesdropper) locations, subject to a maximum threshold on the sensing PCRB. The formulated problem is non-convex and difficult to solve. To deal with this problem, we first show that the problem can be solved via a \emph{two-stage} method, by first obtaining the optimal beamforming corresponding to any given threshold on the signal-to-interference-plus-noise ratio (SINR) at the eavesdropper, and then obtaining the optimal threshold and consequently the optimal beamforming via one-dimensional search of the threshold. By applying the Charnes-Cooper equivalent transformation and semi-definite relaxation (SDR) technique, we relax the first problem into a convex form and further prove that the rank-one relaxation is \emph{tight}, based on which the \emph{optimal solution} of the original beamforming optimization problem can be obtained via the two-stage method with polynomial-time complexity. Then, we further propose two suboptimal solutions with lower complexity by designing the information beam and/or AN beams in the null spaces of the possible eavesdropper channels and/or user channel, respectively. Numerical results validate the effectiveness of our designs in achieving secure communication and sensing in the challenging scenario with unknown target (eavesdropper) location.
\end{abstract}
\section{Introduction}
The sixth-generation (6G) wireless network is anticipated to empower various new applications that require \emph{sensing} the environment, such as autonomous vehicles and monitoring/surveillance \cite{bibWSaad}. Recently, integrated sensing and communication (ISAC) has attracted significant research attention as it can realize simultaneous sensing and communication in a cost-effective manner. Specifically, ISAC enables the simultaneous utilization of wireless infrastructures and limited spectrum/power resources for both communication and sensing purposes, leading to a paradigm shift in wireless networks \cite{bibFanLISAC,bibALiu,bibMLRahman}. On one hand, ISAC can achieve enhanced sensing performance by leveraging the ubiquitous coverage and connectivity provided by wireless networks \cite{bibQinShi}. On the other hand, the environment information (e.g., locations of scatters) sensed via ISAC facilitates intelligent decision-making and adaptation of communication networks, leading to improved performance, enhanced user experience, and efficient resource utilization \cite{bibXCheng}. 

ISAC typically realizes both communication and sensing functionalities via joint signal processing design at the transmitter. To achieve high-accuracy sensing, the dual-functional signal should generally be designed such that the power focused at the target is high to strike strong echo signals in device-free sensing \cite{bibQinShi}, or to achieve high signal-to-noise ratio (SNR) in device-based sensing \cite{ZZhangICC,KareemICC}. However, since the dual-functional signal also carries desired information for the communication user, this gives rise to a severe security risk if an eavesdropper exists in the vicinity of the target, or the target serves as a \emph{potential eavesdropper} \cite{bibZWeiCommunM,bibAD, bibBKC_SPL, bibSMaTWC, bibNSu_TWC}. Particularly, the base stations (BSs) responsible for sensing a particular target are typically located closely to the target with few or no scatters in the BS-target channels. Thus, the BS-target channels are generally stronger than the channels between the BS and the communication user, which makes secrecy communication more challenging. 

In the literature, several prior works have studied the transmit signal design towards secure communication in ISAC. \cite{bibAD} and\cite{bibZ_conf} proposed that the BS transmit signal can be designed as a combination of information signals and artificial noise (AN) signals to ensure the sensing quality while preventing eavesdropping, where the location parameters to be estimated are assumed to be perfectly known before the beamforming design to facilitate target tracking and detection at a given location. On the other hand, it is worth noting that various works on ensuring communication secrecy via physical-layer security techniques are based on the assumption that the BS-eavesdropper channel is known perfectly at the BS \cite {bibSC1, bibSC2, bibSC3}. 

There has also been a line of works that studied the case where the target's location parameters to be sensed is inaccurately known at the BS with error. For such cases, \cite{bibNSu_2021TWC} studied the joint transmit beamforming design to minimize the sum of the received SNR over a range of possible target locations while guaranteeing a signal-to-interference-plus-noise ratio (SINR) level for the legitimate user; \cite{bibMengHua} studied the robust optimization of the passive beamforming at an intelligent reflecting surface (IRS) and the active beamforming at the BS to maximize the sensing beam gain towards the target subject to minimum SINR constraints at the communication users and a maximum tolerable information leakage constraint to the target. Moreover, \cite{bibZRen} proposed a beampattern matching method to jointly optimize the transmitted confidential information signals with additional dedicated sensing signals to minimize the weighted sum of beampattern matching errors subject to a secure communication constraint. On the other hand, a novel optimization framework was considered in \cite{DXu_scan} that jointly optimizes the communication and sensing resources over a sequence of snapshots based on scanning a sequence slice of the sector. In \cite{bib_NanS_iter}, the sensing and secure communication functionalities were improved iteratively, where the {{Cram\'er-Rao bound} (CRB)} was utilized as the sensing performance metric. Note that the transmit signal optimization was still tailored for {\emph{deterministic}} and {\emph{known}} parameters.

However, in practice, the target's location parameter to be sensed can be \emph{unknown} and \emph{random}, while its  distribution can be known \emph{a priori} based on empirical data or target movement pattern \cite{bibXuChan_arXiv}. By properly exploiting such prior information, effective sensing can be completed in one shot instead of iteratively over multiple time slots. A metric to characterize the sensing performance exploiting such prior distribution information is the {\it{posterior Cram\'er-Rao bound} (PCRB)} or Bayesian Cram\'er-Rao bound (BCRB) \cite{bibTrees}, which quantifies a global lower bound for the mean-squared error (MSE) of unbiased estimators. Along this line, \cite{bibXuChan_arXiv,Xu_Journal} derived the PCRB when the location parameter of a target is estimated via a multiple-input multiple-output (MIMO) radar system exploiting its prior distribution information. Based on this, the transmit signal optimization problem to minimize the PCRB was studied, where the proposed solution was observed to focus more power to the locations with high probabilities, and less power to the locations with small probabilities. Nevertheless, for a general ISAC system where the target can potentially serve as an eavesdropper, how to design the transmit beamforming to achieve secure communication and high-quality sensing via exploiting target location distribution still remains an open problem with unique new challenges: \emph{i)} to enhance the sensing performance, more power needs to be focused on the highly-probable target locations, while to avoid information leakage, less power should be focused over the possible target locations, which results in a non-trivial conflict in the transmit beamforming optimization; \emph{ii)} without the exact eavesdropper location, the channel from the BS to the eavesdropper is unknown, which calls for new statistical measures of the communication secrecy for optimizing the transmit beamforming.

\begin{figure}
\centering
\includegraphics[width=9cm]{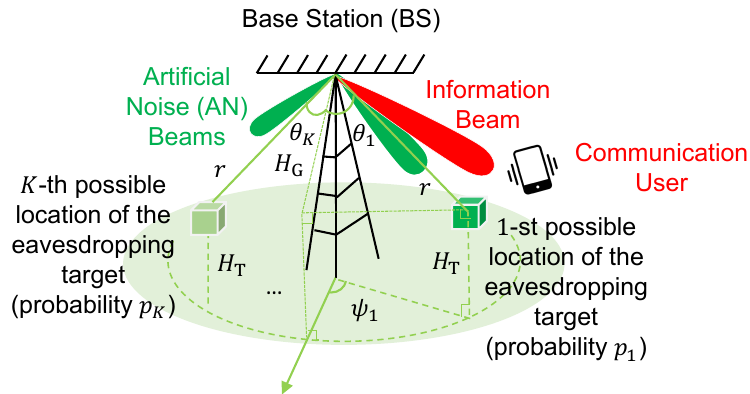}
\caption{Illustration of a secure ISAC system with random target (eavesdropper) location.}\label{Fig1_system}
\end{figure}

Motivated by the above, this paper studies a multi-antenna ISAC system with one multi-antenna BS that aims to communicate with a single-antenna user and sense the location parameter of a target via the echo signals reflected by the target and received back at the BS receive antennas, as illustrated in Fig. \ref{Fig1_system}. The location parameter of the target is \emph{unknown} and \emph{random}, while its distribution information is known \emph{a priori} for exploitation. Specifically, the target has $K\geq 1$ possible locations, for which the probability mass function (PMF) is known. The target also serves as a potential eavesdropper of the communication user. Our main contributions are summarized as follows.
\begin{itemize}
\item First, to systematically characterize the sensing performance, we propose to approximate the discrete PMF with a differentiable probability density function (PDF) under a Gaussian mixture model, so as to adopt the PCRB framework which is only suitable for differentiable probability distributions. Specifically, the approximated PDF is the weighted superposition of multiple Gaussian PDFs, each with mean being a possible target location parameter, variance being a small value, and weight being the probability mass for the corresponding location. Based on this model, we derive the exact PCRB, which is in a complicated form of the transmit covariance matrix. To draw more insights, we propose an approximation of the PCRB in \emph{closed form}, which is observed to be tight numerically. It can be shown that to minimize the approximated PCRB, the transmit signal design should realize a \emph{``probability-dependent power focusing''} effect, where the amount of power focused over each location increases with its associated probability.
\item Next, we propose to adopt an AN-based transmit beamforming structure, where the transmitted signal vector at the BS is the superposition of an information beam and multiple AN beams. The motivation for this approach lies in two aspects: \emph{i)} the AN beams can increase the design flexibility in achieving probability-dependent power focusing over multiple possible locations, for which the degree-of-freedom (DoF) provided by only the information beam to the single-antenna user is generally limited and insufficient; \emph{ii)} the AN beams can focus power over possible target (eavesdropper) locations without leaking information, and create additional noise at the eavesdropper, thus enhancing the secrecy of communication.

\item Under the AN-based transmit beamforming structure, we formulate a new beamforming optimization problem for both the information beam and AN beams at the transmitter to maximize the \emph{worst-case} secrecy rate among all possible target locations, subject to a constraint on the sensing PCRB. The problem is non-convex and difficult to solve due to the newly introduced worst-case secrecy rate and PCRB constraint. To tackle this problem, we first show that the problem can be solved via a two-stage method, where the optimal beamforming corresponding to any given threshold on the SINR at the eavesdropper is obtained first, and the optimal threshold and consequently the globally optimal beamforming is then obtained via one-dimensional search of the threshold. To solve the first problem, we apply the semi-definite relaxation (SDR) technique together with the Schur complement technique and Charnes-Cooper transformation to relax the problem into a convex semi-definite program (SDP), and prove that the rank-one relaxation is \emph{tight}. Based on this, the \emph{optimal solution} to the original beamforming optimization problem can be obtained in \emph{polynomial time}. 

\item Furthermore, we propose two suboptimal solutions with lower computational complexities. Both suboptimal solutions constrain the AN beams to reside within the null space of the communication user channel such that there is no interference to the user; while two different information beam design approaches are considered. For the first one, the information beam is restricted to lie in the null space of the channels for all possible target locations to avoid information leakage; while for the second one, the information beam is aligned to the user channel to maximize the desired signal power at the user. Efficient optimization algorithms are proposed to optimize the beamforming design under these structures, for which the complexities are analyzed to be lower than that of the optimal solution.

\item Finally, we provide numerical results to evaluate the performance of our proposed designs. It is observed that the proposed algorithm is effective in finding the optimal solution. With the optimal solution, the amount of power focused over the possible target locations generally reaches its local minimum for the information beam, and its local maximum for the AN beams. Particularly, the power of the AN beams focused on each possible location generally increases with its probability. Moreover, there exists a non-trivial trade-off between the secrecy rate and the sensing PCRB threshold. Furthermore, the optimal solution outperforms the two suboptimal solutions, at the cost of higher computational complexity.
\end{itemize}

The rest of this paper is organized as follows. Section \ref{Section Systemmodel} introduces the system model and proposes the AN-based transmit beamforming structure. Section \ref{SectionSensingPCRB} presents the Gaussian mixture model to approximate the PMF, derives the PCRB, and proposes a tight approximation of it to draw insights. Then, the joint information and AN beamforming optimization problem is formulated in Section \ref{SectionPF}. Section \ref{SectionPS} presents the optimal solution and suboptimal solutions to the formulated problem, for which the computational complexities are also analyzed. Section \ref{SectionSimulation} provides numerical results. Finally, Section \ref{SectionConclusion} concludes this paper.

\textit{Notations:} Boldface upper-case letters and boldface lower-case letters denote matrices and vectors, respectively. $\mathbb{C}^{N\times L}$ denotes the space of $N\times L$ complex matrices. $\mathbb{R}^{N\times L}$ denotes the space of $N\times L$ real matrices. $\mv{I}_{N}$ denotes an $N\times N$ identity metrix, and $\mv{0}$ denotes an all-zero matrix with appropriate dimension. $j = \sqrt{-1}$ denotes the imaginary unit. $(\cdot)^T$ denotes the transpose operation, and $(\cdot)^H$ denotes the conjugate transpose operation. For a complex number, $|\cdot|$, $(\cdot)^{*}$, and ${\mathfrak{Re}}\left\{\cdot\right\}$ denote the absolute value, conjugate value, and real part, respectively. For a vector, $\|\cdot\|$ denotes its $l_2$-norm. For an arbitrary-sized matrix, ${\rm{rank}}(\cdot)$ and $[\cdot]_{i,j}$ denote its rank and $(i,j)$-th element, respectively. ${\rm{diag}}\{x_1,...,x_M\}$ denotes an $M\times M$ diagonal matrix with $x_1,...,x_M$ being the diagonal elements. For a square matrix, ${\rm{det}}(\cdot)$, ${\rm{tr}}(\cdot)$, and $(\cdot)^{-1}$ denote its determinant, trace, and inverse, respectively. $\mv{X}\succeq \mv{0}$ means that $\mv{X}$ is a positive semi-definite matrix. $\mathcal{CN}(\mu,\sigma^2)$ denotes the distribution of a circularly symmetric complex Gaussian (CSCG) random variable with mean $\mu$ and variance $\sigma^2$, where $\sim$ denotes ``distributed as''. $\mathbb{E}[\cdot]$ denotes the statistical expectation. $[a]^+{=}\max\{a,0\}$ denotes the maximum between a real number $a$ and $0$. $\mathcal{O}(\cdot)$ denotes the standard big-O notation. For a function $\mv{f}(x)$, $\dot{\mv{f}}(x)=\frac{\partial \mv{f}(x)}{\partial x}$ 

\section{System Model} \label{Section Systemmodel}
We consider a secure multi-antenna ISAC system where a BS equipped with $N_t\geq 1$ transmit antennas and $N_r\geq 1$ co-located receive antennas serves a single-antenna communication user in the downlink. Both the transmit and receive antennas at the BS follow a uniform linear array (ULA) layout. Moreover, the BS aims to sense the location parameter of a target which serves as a potential \emph{eavesdropper} based on the received echo signals reflected by the target.\footnote{Note that although the target may serve as an eavesdropper and potentially possess signal reception/transmission capability that enables active sensing\cite{ZZhangICC}, it does not have an incentive to proactively share its location with the BS. Thus, we consider device-free passive sensing via echo signals.} The exact location information of the target is \emph{unknown} and \emph{random}, while its distribution is available to be exploited as prior information. Specifically, the target has $K\geq 1$ possible locations denoted by $\mathcal{K}=\{1,...,K\}$. Under a three-dimensional (3D) coordinate system with the reference point of the BS being the origin, as illustrated in Fig. \ref{Fig1_system}, we consider a fixed height of the eavesdropper denoted by $H_{\mathrm{T}}$ in meters (m), while the height of the BS antennas is denoted as $H_{\mathrm{G}}$ m. Moreover, for ease of revealing fundamental insights, we consider a common distance from the BS to each $k$-th possible target location denoted by $r$ m, which is assumed to be known \emph{a priori},\footnote{The range information $r$ can be obtained by exploiting empirical observations or estimated \emph{a priori} using e.g., time-of-arrival (ToA) methods.} and a different azimuth angle denoted by $\psi_k\in \left[-\frac{\pi}{2},\frac{\pi}{2}\right)$.\footnote{Note that our results can be extended to the case with uniform planar array (UPA) at the BS, where the BS can sense a wider range of angles in $[-\pi,\pi)$ without ambiguity.} The angle-of-departure (AoD) of the signal transmitted from the BS to the $k$-th possible target location, and the angle-of-arrival (AoA) of the signal reflected from the $k$-th possible target location to the BS are then given by
\begin{align}
\!\!\!\theta_k\!=\!{\rm{arcsin}}\!\left(\sin \psi_k\frac{H_{\mathrm{G}}-H_{\mathrm{T}}}{r}\!\right)\!\!\in \!\!\left[-\frac{\pi}{2},\frac{\pi}{2}\right)\!,~ k\in \mathcal{K}.
\end{align}
In this paper, we aim to sense the aforementioned AoD/AoA corresponding to the target, which is termed as the ``\emph{angle}'' of the target for brevity and denoted by $\theta\in \left[-\frac{\pi}{2},\frac{\pi}{2}\right)$. Let $p_k\in [0,1]$ denote the probability for the target to appear at the $k$-th location, with $\sum_{k=1}^K p_k=1$. The PMF of $\theta$ is thus given by
\begin{align}
p_{\Theta}(\theta)=
\begin{cases}
	p_k,&{\rm{if}}~\theta=\theta_k,~k\in \mathcal{K}, \\ 
	0, & {\rm{otherwise.}}\label{cons: p_theta_dis}
\end{cases}
\end{align}
We consider a challenging scenario for secrecy communication where the target has a (strong) line-of-sight (LoS) channel with the BS, and the downlink eavesdropping channel denoted by $\mv{h}_{\mathrm{E}}^H(\theta)\in \mathbb{C}^{1\times N_t}$ is \emph{unknown} due to the unknown target's angle $\theta$. On the other hand, the channel from the BS to the user denoted by $\mv{h}^H\in \mathbb{C}^{1\times N_t}$ is assumed to be perfectly known at the BS. Furthermore, we assume $\mv{h}_{\mathrm{E}}^H(\theta_k)$ for $k\in \mathcal{K}$ and $\mv{h}^H$ are linearly independent, which can hold for various user channel models including the LoS model (with a distinctive user angle) and random Rayleigh fading model.

We aim to optimize the BS transmit signals via smart exploitation of the \emph{distribution information} about the eavesdropping target's angle to achieve secure communication and also accurately sense the angle of the eavesdropping target.\footnote{Note that the sensing result may facilitate more tailored signal designs with further improved secrecy performance in future channel uses, while the joint signal design over multiple channel uses is left as our future work.} To this end, the unknown and random angle of the eavesdropping target brings two new challenges. Firstly, the secrecy communication rate becomes random, which calls for new statistical performance metrics. Secondly, to accurately sense the target's angle, the signal needs to be beamed towards multiple possible angles of the target based on their probabilities to ``statistically'' strengthen the echo signal reflected by the target. 

Motivated by the above, in this paper, we aim to maximize the worst-case secrecy rate corresponding to the most favorable location for eavesdropping, under a sensing accuracy constraint. We further introduce an {\emph{AN-based beamforming design}}, where the transmitted signal vector is the superposition of an information beam and $J\leq N_t$ AN beams. Denote $s\sim \mathcal{CN}(0,1)$ as the information symbol for the user, and $\mv{w} \in \mathbb{C}^{N_t\times 1}$ as the information beamforming vector. We further denote $\mv{v}_j\in \mathbb{C}^{N_t\times 1} $ as the $j$-th AN beamforming vector, and $s_j\sim\mathcal{CN}(0,1)$ as the $j$-th independent AN signal which is also independent of $s$. The transmitted signal vector is thus given by
\begin{align} \label{transmitted signal x}
\mv{x} = \mv{w}s+\sum_{j=1}^{J}\mv{v}_js_j.
\end{align}
The transmit covariance matrix is consequently given by $\mv{R}_x=\mathbb{E}[\mv{x}\mv{x}^H]=\mv{ww}^H+\sum_{j=1}^{J}\mv{v}_j\mv{v}_j^H$. Let $P$ denote the transmit power constraint, which yields $\mathbb{E}[\|\mv{x}\|^2]=\|\mv{w}\|^2+\sum_{j=1}^{J}\|\mv{v}_j\|^2\leq P$ and equivalently $\mathrm{tr}(\mv{R}_x)\leq P$. Note that the motivation for the AN-based approach is two-fold. Firstly, by introducing additional Gaussian-distributed noise signals which are the worst-case noise for eavesdropping, the received SINR at the potential eavesdropper can be decreased, thus enhancing the \emph{communication secrecy}. Secondly, the extra AN beams provide more design flexibility in strengthening the echo signals reflected from multiple possible target angles, thus enhancing the \emph{sensing accuracy}.

Based on (\ref{transmitted signal x}), the received signal at the user is given by 
\begin{align} 
y = \mv{h}^H\mv{x} + z = \mv{h}^H\mv{w}s+\mv{h}^H\sum_{j = 1}^{J}\mv{v}_js_j+z,
\end{align}
where $z\sim\mathcal{C}\mathcal{N}\left( {0,\sigma^{2}}\right)$ denotes the CSCG noise at the user receiver with average power $\sigma^2$. The SINR at the user receiver is thus given by
\begin{equation}
{\rm{SINR}} = \frac{| {\mv{h}^{H}\mv{w}} |^{2}}{\sum_{j=1}^J|{\mv{h}^{H}\mv{v}_j} |^{2} + \sigma^{2}}.\label{con: SINR CU}	\vspace{-2mm}
\end{equation}
The received signal at the eavesdropping target is given by
\begin{equation}
\!\!\!	{y_\mathrm{E}}({\theta})\! =\! {\mv{h}}_\mathrm{E}^H(\theta){\mv{x}}\!+\!z_{\mathrm{E}}\!=\!{\mv{h}}_{\mathrm{E}}^H(\theta){\mv{w}}s\!+\!{\mv{h}}_{\mathrm{E}}^H(\theta)\sum_{j=1}^J\mv{v}_js_j\!+\!z_{\mathrm{E}},\!\!\! \label{con: RsignalEaves}
\end{equation}
where $z_\mathrm{E}\!\sim\!\mathcal{C}\mathcal{N}({0,\sigma_{\mathrm{E}}^{2}})$ denotes the CSCG noise at the eavesdropper receiver with $\sigma^2_{\mathrm{E}}$ denoting the average noise power. Specifically, under the LoS BS-eavesdropper channel and the ULA layout assuming half wavelength antenna spacing, we have ${\mv{h}}_\mathrm{E}^H(\theta)=\frac{\sqrt{\beta_0}}{r}{\mv{a}}^H(\theta)$, where $\beta_0$ denotes the reference channel power at 1 m; ${\mv{a}}^H(\theta)=[e^{- j\frac{\pi(N_{t} - 1){\sin\theta}}{2}}, e^{-j\frac{\pi(N_{t} - 3){\sin\theta}}{2}},...,e^{j\frac{\pi(N_{t} - 1){\sin\theta}}{2}}]$ denotes the steering vector at the BS transmit array. Hence, the SINR at the potential eavesdropper can be expressed as
\begin{align}
{\rm{SINR}}_{\mathrm{E}}(\theta)\! =\! \frac{| {\mv{a}^{H}(\theta)\mv{w}} |^{2}}{\sum_{j=1}^J| {\mv{a}^{H}(\theta)}\mv{v}_j |^{2}\! +\! \frac{\sigma_{\mathrm{E}}^2r^2}{\beta_0}},\theta\!\in\! \{\theta_1,...,\theta_K\}.
\end{align}
The achievable secrecy rate at the user when there exists an eavesdropper at location $k$ with angle $\theta_k$ is given by \cite{bib9}:
\begin{align}\label{cons: r_0}
\!\!\!\!R_k\!=[\log_2(1\!+\!\mathrm{SINR})\!-\!\log_2(1\!+\!\mathrm{SINR}_{\mathrm{E}}(\theta_k))]^+,k\in \mathcal{K}\!\!
\end{align}
in bps/Hz. The worst-case achievable secrecy rate among all possible eavesdropper locations is thus given by $R=\underset{k\in \mathcal{K}}{\min}\ R_k$.

On the other hand, the transmit signal will be reflected by the target. Let $\alpha\in \mathbb{C}$ denote the radar cross section (RCS) coefficient of the target, which is generally an \emph{unknown} and \emph{deterministic} parameter.\footnote{It is worth noting that our results are also applicable to the case where the RCS coefficient is random with an unknown distribution.} Let ${\mv{b}}(\theta)\!=\![e^{-j\frac{\pi(N_{r}-1)\sin\theta}{2}}, e^{-j\frac{\pi(N_{r}-3)\sin\theta}{2}},...,e^{j\frac{\pi(N_{r}-1)\sin\theta}{2}}]^H$ denote the steering vector at the BS receive array. The received echo signal at the BS receive antennas is given by
\begin{align} \label{cons: y_R}
\mv{y}_{\mathrm{R}} = \frac{\beta_0}{r^2}\mv{b}(\theta)\alpha\mv{a}^{H}(\theta)\mv{x} + \mv{z}_{\mathrm{R}} \triangleq\beta 
\mv{M}(\theta)\mv{x}+\mv{z}_{\mathrm{R}},
\end{align}
where $\beta\overset{\Delta}{=}\frac{\beta_0}{r^2}\alpha$ denotes the overall reflection coefficient including the two-way channel and RCS coefficient; ${\mv{z}}_{\mathrm{R}}\!\sim\!\mathcal{CN}(\mv{0},\sigma_{\mathrm{R}}^2{\mv{I}}_{N_r})$ denotes the CSCG noise vector with $\sigma_{\mathrm{R}}^2$ denoting the average noise power at the BS receive antennas; and ${\mv{M}}(\theta)\overset{\Delta}{=}\mv{b}(\theta)\mv{a}^{H}(\theta)$.

Note that both the worst-case secrecy rate $R$ and the received echo signal in $\mv{y}_{\mathrm{R}}$ are determined by the transmit signal vector $\mv{x}$ and consequently the beamforming vectors $\mv{w}$ and $\{\mv{v}_j\}_{j=1}^J$. To formulate the beamforming optimization problem, our remaining task is to characterize the performance of estimating (sensing) $\theta$ based on the received signal vector in (\ref{cons: y_R}), which will be addressed in the next section. 

\section{Sensing Performance Characterization Exploiting Target Location Distribution}\label{SectionSensingPCRB}
Notice from (\ref{cons: y_R}) that the overall reflection coefficient $\beta=\beta_R+j\beta_I$ is also an unknown (and deterministic) parameter, which thus also needs to be estimated to obtain an accurate estimation of $\theta$. Let ${\mv{\omega}}=[\theta,\beta_R,\beta_I]^T$ denote the collection of unknown parameters to be estimated. With the prior distribution information of $\theta$ available for exploitation, we propose to employ \emph{PCRB} as the performance metric, which characterizes a global lower bound of the MSE of unbiased estimators exploiting prior information. To the best of our knowledge, most classic PCRB derivation methods are suitable for estimation parameters with continuous and differentiable PDFs \cite{bibTrees}. For consistency, we propose to approximate the discrete PMF in (\ref{cons: p_theta_dis}) with a continuous Gaussian mixture PDF given by
\vspace{-1mm}\begin{equation}
\bar{p}_\Theta(\theta)= \sum_{k=1}^{K}p_k\frac{1}{\sqrt{2\pi}\sigma_{\theta}}e^{-\frac{(\theta-\theta_k)^2}{2\sigma_{\theta}^2}}. \label{cons: barptheta}	\vspace{-1mm}
\end{equation}
Specifically, $\bar{p}_\Theta(\theta)$ is the weighted sum of $K$ Gaussian PDFs, where each $k$-th Gaussian PDF is centered at mean $\theta_k$ with a small variance $\sigma_\theta^2$, and carries a weight of $p_k$. Note that as $\sigma_\theta^2$ decreases, $\bar{p}_\Theta(\theta)$ becomes increasingly similar to ${p}_\Theta(\theta)$. Moreover, with a sufficiently small $\sigma_\theta^2$, the probability for $\theta$ under (\ref{cons: barptheta}) to exceed the original $\left[-\frac{\pi}{2},\frac{\pi}{2}\right)$ region is negligible.

Based on (\ref{cons: barptheta}), the Fisher information matrix (FIM) for the estimation of $\mv{\omega}$ consists of two parts as follows \cite{bib_Shen}:
\vspace{-2mm}\begin{equation}
\mv{F} = \mv{F}_{\mathrm{D}} + \mv{F}_{\mathrm{P}}.	\vspace{-2mm}
\end{equation}
The first part $\mv{F}_{\mathrm{D}}\in \mathbb{R}^{3\times 3}$ represents the FIM extracted from the observed data in ${\mv{y}}_{\mathrm{R}}$, which is given by
\begin{align}
\left\lbrack \mv{F}_{\mathrm{D}} \right\rbrack_{i,j}=-\mathbb{E}_{\mv{y}_{\mathrm{R}},\mv{\omega}}\left\lbrack \frac{\partial^{2}L_{\mv{y}_{\mathrm{R}}}(\mv{\omega})}{\partial\omega_{i}\partial\omega_{j}} \right\rbrack,\ i,j\in \{1,2,3\},
\end{align}
with $L_{\mv{y}_{\mathrm{R}}}(\mv{\omega})\!=\!-N_r{\rm{ln}}(\pi\sigma_{\mathrm{R}}^2)\!-\!\frac{1}{\sigma_{\mathrm{R}}^2}(\|\mv{y}_{\mathrm{R}}\|^2\!+\!|\beta|^2\|\mv{M}(\theta)\mv{x}\|^2)+\frac{2}{\sigma_{\mathrm{R}}^2}\mathfrak{Re}\{\beta^*\mv{x}^H\mv{M}^H(\theta)\mv{y}_{\mathrm{R}}\}$ being the log-likelihood function for the parameters in $\mv{\omega}$. $\mv{F}_{\mathrm{D}}$ can be further derived as
\begin{align}
\mv{F}_{\mathrm{D}} = \begin{bmatrix}
	F_{\theta\theta}&\mv{F}_{\theta\beta} \\ 
	\mv{F}_{\theta\beta}^H&  \mv{F}_{\beta\beta}
\end{bmatrix},
\end{align}
where each block is given as
\begin{align}
\!\!\!&F_{\theta\theta}= \frac{2|\beta|^2}{\sigma_{\mathrm{R}}^{2}}\int_{-\infty}^{\infty}\bar{p}_\Theta(\theta){{\rm{tr}( {\dot{\mv{M}}^{\it{H}}( \theta )\dot{{\mv{M}}}( \theta )\mv{R}_{\mathit{x}}} )}}{\rm{d}}\theta,\\
&\mv{F}_{\theta\beta}= 
\!\!\frac{2}{\sigma_{\mathrm{R}}^{2}}\!\!\int_{-\infty}^{\infty}\!\!\bar{p}_\Theta(\theta)\mathfrak{Re}\{\beta^*{\rm{tr}}(\dot{\mv{M}}^H(\theta)\mv{M}(\theta)\mv{R}_x)[1,j]\}
{\rm{d}}\theta,\\
&\mv{F}_{\beta\beta}= \frac{2}{\sigma_{\mathrm{R}}^{2}}\int_{-\infty}^{\infty}\bar{p}_\Theta(\theta){\rm{tr}}( \mv{M}^{H}( \theta )\mv{M}( \theta )\mv{R}_{\mathit{x}} )\mv{I}_{2}{\rm{d}}\theta.
\end{align}

The second part $\mv{F}_{\mathrm{P}}\in \mathbb{R}^{3\times 3}$ represents the FIM extracted from the prior distribution information, which is given by
\begin{align}
\left\lbrack \mv{F}_{\mathrm{P}} \right\rbrack_{i,j} = - \mathbb{E}_{\mv{\omega}}\left\lbrack \frac{\partial^{2}{\ln p_{\mv{w}}}( \mv{\omega} )}{\partial\omega_{i}\partial\omega_{j}} \right\rbrack,\ i,j\in \{1,2,3\},
\end{align}
where $p_{\mv{w}}(\mv{\omega})$ denotes the PDF of $\mv{\omega}$. Note that since $\beta_R$ and $\beta_I$ are both deterministic variables, $\mv{F}_{\mathrm{P}}$ only has a non-zero entry in the first column and first row, which is given by
\begin{align} \label{cons: J_D intergral}
\left\lbrack \mv{F}_{\mathrm{P}} \right\rbrack_{\theta\theta} = -\int_{\infty}^{\infty}\frac{\partial^2\bar{p}_\Theta(\theta)}{\partial^2 \theta}d\theta+ \int_{\infty}^{\infty}\frac{\left(\frac{\partial  \bar{p}_\Theta(\theta)}{\partial \theta}\right)^2}{\bar{p}_\Theta(\theta)}d\theta. 
\end{align}
The first term in the right-hand side of (\ref{cons: J_D intergral}) can be derived as
\begin{align}
\!\!\!-\int_{-\infty}^{\infty}\frac{\partial^2\bar{p}_\Theta(\theta)}{\partial^2 \theta}d\theta\!= \!\!\sum_{k=1}^{K}\frac{p_k(\theta-\theta_k)}{\sigma_\theta^3\sqrt{2\pi}}{\rm{e}}^{-\frac{(\theta-\theta_k)^2}{2\sigma_\theta^2}}\bigg|_{-\infty}^{\infty}\!=\!0.
\end{align}
The second term can be derived as
\begin{align}
\int_{-\infty}^{\infty}\frac{\Big(\frac{\partial  \bar{p}_\Theta(\theta)}{\partial \theta}\Big)^2}{\bar{p}_\Theta(\theta)}d\theta \!=&\!\! \int_{-\infty}^{\infty}\sum_{k=1}^{K}(\theta-\theta_k)^2\Big(\frac{\bar{p}_\Theta(\theta)}{\sigma_{\theta}^4}\Big)d\theta 
-\epsilon \nonumber \\
= &\sum_{k=1}^{K}p_k\frac{1}{\sigma^2_\theta}-\epsilon= \frac{1}{\sigma^2_\theta}-\epsilon,
\end{align}
where $\epsilon\!\overset{\Delta}{=}\!\int_{-\infty}^\infty\sum\limits_{k=1}^{K}\!\sum\limits_{n=1 }^{K}f_{k}(\theta)f_{n}(\theta)\frac{(\theta_n-\theta_{k})^2}{\sigma_{\theta}^2}/(2\sum\limits_{k=1}^{K}f_k(\theta))d\theta$ with $f_k(\theta)\!\!\overset{\Delta}{=}\!\!\frac{p_k}{\sqrt{2\pi}\sigma_{\theta}}e^{-\frac{(\theta\!-\!\theta_k)^2}{2\sigma_{\theta}^2}}$. Thus, we have $\left\lbrack \mv{F}_{\mathrm{P}} \right\rbrack_{\theta\theta}\! =\! \frac{1}{\sigma^2_\theta}-\epsilon$. 

Based on the above, the overall FIM $\mv{F}$ is given by
\begin{align}
\mv{F} = \mv{F}_{\mathrm{D}}+\mv{F}_{\mathrm{P}} = \begin{bmatrix}
	F_{\theta\theta}+\frac{1}{\sigma^2_\theta}-\epsilon&\mv{F}_{\theta\beta} \\ 
	\mv{F}_{\theta\beta}^H &  \mv{F}_{\beta\beta}
\end{bmatrix}.
\end{align}
The PCRB for the estimation MSE of ${\theta}$ denoted by ${\rm{PCRB}}_\theta$ is then given by the entry in the first column and first row of $\mv{F}^{-1}$, which can be expressed as
\begin{align}\label{cons:pcrb}
&{\rm{PCRB}}_\theta \!= \!\left[\mv{F}^{-1}\right]_{1,1}\!\!
= \!\left[F_{{\theta}{\theta}}\!+\!\frac{1}{\sigma^2_\theta}\!-\!\epsilon-\mv{F}_{{\theta}{\beta}}\mv{F}_{\beta\beta}^{-1}\mv{F}_{\theta\beta}^{H}\right]^{-1}\!\!\!\\ 
&=\frac{\sigma_{\mathrm{R}}^2{g}_1(\mv{R}_x)}{2|\beta|^2\left(\left({g}_2(\mv{R}_x)+\frac{\sigma_{\mathrm{R}}^2}{2|\beta|^2}\left(\frac{1}{\sigma_\theta^2}-\epsilon\right)\right){g}_1(\mv{R}_x)-\left|{g}_3(\mv{R}_x)\right|^2\right)},\nonumber
\end{align}
where 
\begin{align}
{g}_1(\mv{R}_x)=\int_{-\infty}^{\infty}\bar{p}_\Theta(\theta){\rm{tr}}({{\mv{M}}^{H}(\theta)\mv{M}(\theta )\mv{R}_{\mathit{x}}}){\rm{d}}\theta,\label{g1}\\
{g}_2(\mv{R}_x) =  \int_{-\infty}^{\infty}\bar{p}_\Theta(\theta){\rm{tr}}({\dot{\mv{M}}^{\it{H}}( \theta )\dot{{\mv{M}}}(\theta)\mv{R}_{\mathit{x}}}){\rm{d}}\theta,\label{g2}\\
{g}_3(\mv{R}_x)= \int_{-\infty}^{\infty}\bar{p}_\Theta(\theta){\rm{tr}}({\dot{{\mv{M}}}^{H}(\theta)\mv{M}(\theta)\mv{R}_{\mathit{x}}}){\rm{d}}\theta.\label{g3}
\end{align}

Due to the symmetry in $\mv{a}(\theta)$ and $\mv{b}(\theta)$, we have $\mv{a}^H(\theta)\dot{\mv{a}}(\theta) = 0$ and $\mv{b}^H(\theta)\dot{\mv{b}}(\theta) = 0$. By further noting $\|\mv{b}(\theta)\|^2  = N_r,\forall \theta$, we can further simplify (\ref{g1})-(\ref{g3}) as ${g}_1(\mv{R}_x)=\mathrm{tr}(\mv{M}_1\mv{R}_x)$, ${g}_2(\mv{R}_x)=\mathrm{tr}(\mv{M}_2\mv{R}_x)$, and ${g}_3(\mv{R}_x) = \mathrm{tr}(\mv{M}_3\mv{R}_x)$, where $\mv{M}_1\!\!\! = \!\!\!N_r\int_{-\infty}^{\infty}\bar{p}_\Theta(\theta)\mv{a}(\theta)\mv{a}^H(\theta){\rm{d}}\theta,~\mv{M}_2 = \int_{-\infty}^{\infty}\!\bar{p}_\Theta(\theta)$ $\|\dot{\mv{b}}(\theta)\|^2\mv{a}(\theta)\mv{a}^H(\theta){\rm{d}}\theta\!+\!N_r\int_{-\infty}^{\infty}\!\bar{p}_\Theta(\theta)\dot{\mv{a}}(\theta)\dot{\mv{a}}^H(\theta){\rm{d}}\theta$, and $\mv{M}_3 = N_r\int_{-\infty}^{\infty}\bar{p}_\Theta(\theta)\dot{\mv{a}}(\theta){\mv{a}}^H(\theta){\rm{d}}\theta$. Thus, $\rm{PCRB}_\theta$ can be expressed as
\begin{align} \label{PCRB_theta}
&{\rm{PCRB}}_\theta=\frac{1}{\left(\frac{1}{\sigma_\theta^2}-\epsilon\right)+\frac{2|\beta|^2}{\sigma_{\mathrm{R}}^2}\left(\mathrm{tr}(\mv{M}_2\mv{R}_x)-\frac{|\mathrm{tr}(\mv{M}_3\mv{R}_x)|^2}{\mathrm{tr}(\mv{M}_1\mv{R}_x)}\right)}\nonumber\\
=&1/\Bigg(\Bigg(\frac{1}{\sigma_\theta^2}-\epsilon\Bigg)+\frac{2|\beta|^2}{\sigma_{\mathrm{R}}^2}\Bigg(\mv{w}^H\mv{M}_2\mv{w}
+\sum_{j = 1}^J\mv{v}_j^H\mv{M}_2\mv{v}_j\nonumber\\
&-\frac{|\mv{w}^H\mv{M}_3\mv{w}+\sum_{j = 1}^J\mv{v}_j^H\mv{M}_3\mv{v}_j|^2}{\mv{w}^H\mv{M}_1\mv{w}+\sum_{j = 1}^J\mv{v}_j^H\mv{M}_1\mv{v}_j}\Bigg)\Bigg).
\end{align}
\begin{remark}
It is worth noting that the PCRB in (\ref{PCRB_theta}) is only dependent on the \emph{overall transmit covariance matrix} $\mv{R}_x$, since the information beam and AN beams play the same role in striking the echo signals. Moreover, the PCRB is a decreasing function of the overall reflection gain $|\beta|$ and consequently the gain of the target's unknown RCS coefficient $\alpha$, i.e., $|\alpha|$.
\end{remark}

Note that the exact PCRB in (\ref{PCRB_theta}) is a complicated function with respect to $\mv{R}_x$ and the beamforming vectors $\mv{w}$ and $\mv{v}_j$'s. To draw clearer insights on the effect of beamforming design on the PCRB, we derive a more tractable upper bound of PCRB as follows. Specifically, we first re-express (\ref{PCRB_theta}) as
\begin{align} \label{PCRB_theta_2}
\!\!\!\!{\rm{PCRB}}_\theta
\!=\! \frac{1}{\!\left(\frac{1}{\sigma_\theta^2}\!-\!\epsilon\right)+\frac{2|\beta|^2}{\sigma_{\mathrm{R}}^2}\left(g_4(\mv{R}_x)\!+\! \frac{g_5(\mv{R}_x)}{g_1(\mv{R_x})}\right)},
\end{align}
where ${g}_4(\mv{R}_x) = \int_{-\infty}^{\infty}\bar{p}_\Theta(\theta)\|\dot{\mv{b}}(\theta)\|^2\mv{a}^H(\theta)\mv{R}_x\mv{a}(\theta){\rm{d}}\theta$; ${g}_5(\mv{R}_x) =  \frac{1}{2}\int_{-\infty}^{\infty}\int_{-\infty}^{\infty}N_t^2|\dot{\mv{a}}^H(\theta_p)\mv{R}_x{\mv{a}}(\theta_q)-{\mv{a}}^H(\theta_p)\mv{R}_x\dot{\mv{a}}(\theta_q)|^2
\bar{p}_\Theta(\theta_p)\bar{p}_\Theta(\theta_q)d\theta_pd\theta_q	\geq 0$. 
By noting that both ${g}_1(\mv{R}_x)$ and ${g}_5(\mv{R}_x)$ are non-negative, an upper bound of ${\rm{PCRB}}_\theta$ can be obtained as
\begin{align}\label{Upp1}
&{\rm{PCRB}}_\theta \leq  {\rm{PCRB}}^{\mathrm{U}}_\theta\overset{\Delta}{=}\frac{1}
{\left(\frac{1}{\sigma_\theta^2}-\epsilon\right)+\frac{2|\beta|^2}{\sigma_{\mathrm{R}}^2}{g_4}(\mv{R}_x)}\nonumber\\
=&	\frac{1}{\left(\frac{1}{\sigma^2_\theta}-\epsilon\right)+\frac{2|\beta|^2}{\sigma_{\mathrm{R}}^2}\left(\mv{w}^H\mv{Q}\mv{w}+\sum_{j=1}^J\mv{v}_j^H\mv{Q}\mv{v}_j\right)}, 
\end{align}
where $\mv{Q}=\int_{-\infty}^{\infty}\bar{p}_{\Theta}(\theta)\|\dot{\mv{b}}(\theta)\|^2\mv{a}(\theta)\mv{a}(\theta)^H{\rm{d}}\theta$.  Note that $\mv{Q}$ and $\epsilon$ in (\ref{Upp1}) still involve complicated integrals over the continuous $\theta$. By leveraging the fact that we consider a small variance $\sigma_\theta^2$ in the Gaussian mixture PDF to approximate the PMF, we propose an approximation of (\ref{Upp1}) in \emph{closed form}.  \begin{proposition}\label{prop_app}
With a small $\sigma_\theta^2$, an approximate expression for the PCRB upper bound ${\rm{PCRB}}^{\mathrm{U}}_\theta$ is given by
\begin{align} \label{cons: PCRB_value}
	\!\!\!\!{\rm{PCRB}}^{\mathrm{U}}_\theta\!\approx\!\tilde{\rm{PCRB}}^{\mathrm{U}}_\theta\!\overset{\Delta}{=}\!{\frac{1}{\frac{2|\beta|^2}{\sigma_{\mathrm{R}}^2}(\mv{w}^H\tilde{\mv{Q}}\mv{w}\!+\!\sum_{j=1}^J\mv{v}_j^H\tilde{\mv{Q}}\mv{v}_j)\!+\!\frac{1}{\sigma_\theta^2}}},
\end{align}
where $\tilde{\mv{Q}}\overset{\Delta}{=}\rho_0\sum_{k=1}^K {p_k({\rm{cos}}(2\theta_k)\!+\!1)}\mv{a}(\theta_k)\mv{a}^{H}(\theta_k)$ with $\rho_0\overset{\Delta}{=}\frac{\sum_{n=1}^{N_r}\pi^2(n\!-\!1)^2}{4}$.
\end{proposition}
\begin{IEEEproof}
Please refer to Appendix \ref{proof_prop_app}.
\end{IEEEproof}

\begin{figure}[t]
	\centering
	\includegraphics[width=9cm]{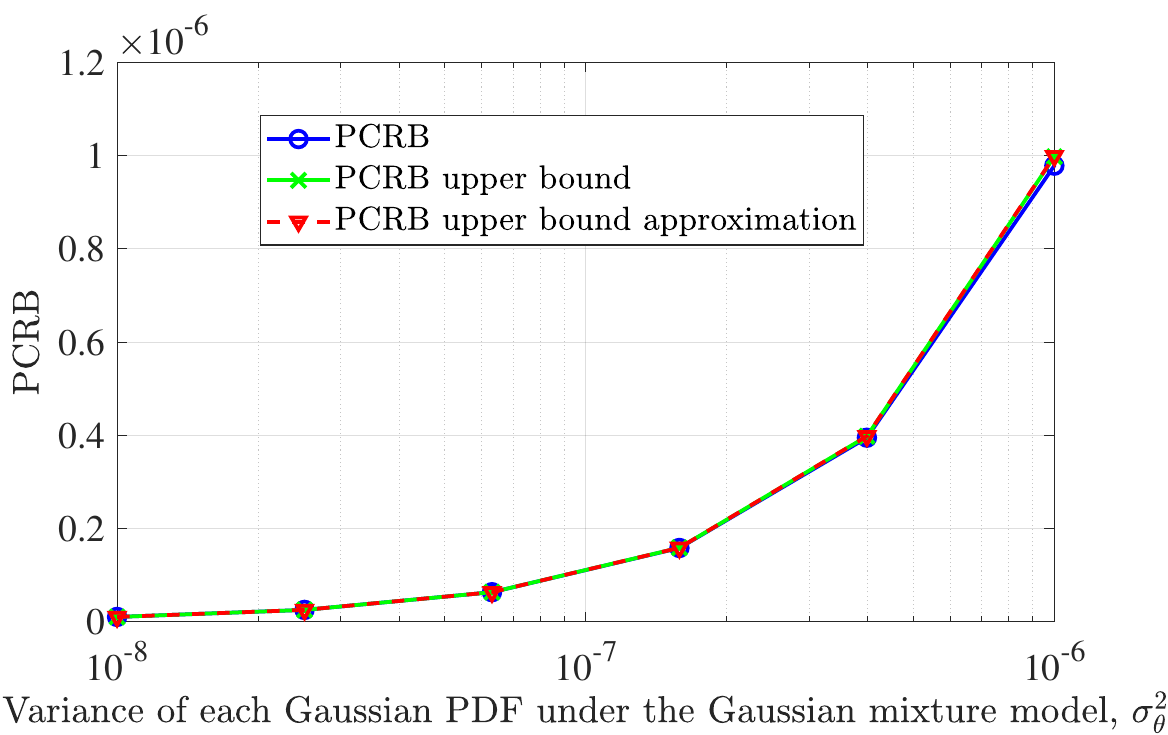}
	\caption{Illustration of the PCRB, PCRB upper bound, and PCRB upper bound approximation.}\label{Bound}
\end{figure}

In Fig. \ref{Bound}, we illustrate the PCRB, PCRB upper bound, and the PCRB upper bound approximation versus $\sigma_\theta^2$, under the same setup as that in Section \ref{SectionSimulation}. It is observed that both the proposed upper bound and the upper bound approximation are close to that of the exact PCRB value.
\begin{remark}[Probability-Dependent Power Focusing]
It can be observed from (\ref{cons: PCRB_value}) that  $\tilde{\rm{PCRB}}^{\mathrm{U}}_\theta$ decreases as the total power of the information beam and AN beams focused on each $k$-th possible target location (which scales with $|\mv{w}^H\mv{a}(\theta_k)|^2+\sum_{j=1}^J|\mv{v}_j^H\mv{a}(\theta_k)|^2$) increases. Moreover, the amounts of power focused on the locations with high probabilities $p_k$'s play more dominant roles in the PCRB. Therefore, to minimize the approximate PCRB upper bound, the transmit beamforming design needs to achieve a so-called \emph{probability-dependent power focusing} effect based on the prior probability distribution.
\end{remark}

Note that optimizing the beamforming to jointly achieve probability-dependent power focusing under sensing performance constraint and secure communication is a non-trivial yet new task, which will be studied in the following sections.

\section{Problem Formulation}\label{SectionPF}
In this paper, our objective is to optimize the transmit beamforming vectors $\mv{w}$ and $\mv{v}_j$'s to maximize the worst-case secrecy rate among all possible eavesdropper locations, while ensuring the sensing PCRB for the eavesdropping target is always below a given threshold $\Gamma$. We aim to achieve this goal by ensuring that the exact PCRB ${\rm{PCRB}}_\theta$ in (\ref{PCRB_theta}) corresponding to the minimum RCS coefficient's gain $|\bar{\alpha}|=\min|\alpha|$ and the resulting $|\bar{\beta}|=\frac{\beta_0}{r^2}|\bar{\alpha}|$ is no larger than $\Gamma$. Thus, the optimization problem is formulated as:
\begin{align} 
({\mbox{P1}})\qquad\qquad&\nonumber\\
{\underset{\mv{w},\{\!\mv{v}_j\!\}_{j=1}^J}{\rm max}}{\underset{k\in \mathcal{K}}{\rm{min}}}\  &\log_2(1\!+\!\mathrm{SINR})\!-\!\log_2(1\!+\!\mathrm{SINR}_{\mathrm{E}}(\theta_k))\!\!\!\!\ \\ 
{\rm{s.t.}} &\ {\left\| \mv{w} \right\|^{2} + \sum_{j=1}^J\left\| \mv{v}_j \right\|^{2} \leq P} \label{P1c1}\\ 
&1/\Bigg(\!\Bigg(\!\frac{1}{\sigma_\theta^2}\!-\!\epsilon\!\!\Bigg)\!\!+\!\!\frac{2|\bar{\beta}|^2}{\sigma_{\mathrm{R}}^2}\!\Bigg(\!\!\mv{w}^H\mv{M}_2\mv{w}
\!+\!\!\sum_{j = 1}^J\mv{v}_j^H\mv{M}_2\mv{v}_j\nonumber\\
&\!-\!\frac{|\mv{w}^H\mv{M}_3\mv{w}\!+\!\sum_{j = 1}^J\mv{v}_j^H\mv{M}_3\mv{v}_j|^2}{\mv{w}^H\mv{M}_1\mv{w}\!+\!\sum_{j = 1}^J\mv{v}_j^H\mv{M}_1\mv{v}_j}\!\Bigg)\!\!\Bigg)\!\!\leq\! \Gamma\!.\!\label{P1c2}
\end{align}

Note that the objective function of Problem (P1) involves logarithm functions of fractional quadratic functions, and can be shown to be non-concave. Moreover, the constraint in (\ref{P1c2}) is also non-convex due to its fractional structure. Therefore, (P1) is a non-convex problem that is challenging to solve. Furthermore, it is worth noting that in order to maximize the secrecy rate, $\mv{w}$ should be designed such that the received power of the information beam at each possible target location is as small as possible, i.e., each possible target location should be an ``information beam hole''; on the other hand, to minimize the sensing PCRB, $\mv{w}$ and $\mv{v}_j$'s should beam the signal towards possible target locations based on their corresponding probabilities (e.g., achieving \emph{probability-dependent power focusing} as discussed in Remark 2). Therefore, there exists a non-trivial trade-off between the secrecy performance and sensing performance in the secure ISAC system. Furthermore, note that both the secrecy rate and the PCRB are critically dependent on the distribution of $\theta$, which makes the problem more challenging. In the following, we solve Problem (P1) via advanced optimization techniques.

\section{Proposed Optimal and Suboptimal Solutions}\label{SectionPS}
\subsection{Feasibility Check of Problem (P1)}\label{sec_feas}
Prior to solving Problem (P1), we check its feasibility. Note that both the power constraint in (\ref{P1c1}) and the PCRB constraint in (\ref{P1c2}) are only dependent on the transmit covariance matrix $\mv{R}_x=\mv{ww}^H+\sum_{j=1}^J \mv{v}_j\mv{v}_j^H$. Thus, the feasibility of (P1) can be checked by solving the following problem.

\begin{align}
\mbox{(P1-F)}\quad {\underset{\mv{R}_x}{\rm max}}\quad  &0 \\ 
{\rm{s.t.}} \quad & \mathrm{tr}(\mv{R}_x) \leq P \\ 
&\mathrm{tr}(\mv{M}_2\mv{R}_x)-\frac{|\mathrm{tr}(\mv{M}_3\mv{R}_x)|^2}{\mathrm{tr}(\mv{M}_1\mv{R}_x)}\geq \xi\label{P1Fc2}\\
&\mv{R}_x\succeq \mv{0},
\end{align}
where $\xi\overset{\Delta}{=}\frac{\sigma_{\mathrm{R}}^2}{2|\bar{\beta}|^2}\left(\frac{1}{\Gamma}\!-\!\frac{1}{\sigma^2_\theta}+\epsilon\right)$.

Note that (\ref{P1Fc2}) involves a complicated fractional expression, which is further simplified by leveraging the Schur complement. Note that ${\rm{tr}}(\mv{M}_1\mv{R}_x)> 0$ holds as long as $\mv{R}_x\neq \mv{0}$, thus ${\rm{tr}}\left(\mv{M}_2\mv{R}_x\right)-\frac{\left|{\rm{tr}}\left(\mv{M}_3\mv{R}_x\right)\right|^2}{{\rm{tr}}\left(\mv{M}_1\mv{R}_x\right)}$ is the Schur complement of  ${\rm{tr}}(\mv{M}_1\mv{R}_x)$ in matrix $\begin{bmatrix}
{\rm{tr}}(\mv{M}_2\mv{R}_x) & {\rm{tr}}(\mv{M}_3\mv{R}_x)\\
{\rm{tr}}(\mv{M}_3^H\mv{R}_x)& {\rm{tr}}(\mv{M}_1\mv{R}_x)
\end{bmatrix}$. According to the Schur complement condition \cite{Schur}, Problem (P1-F) is equivalent to
\begin{align} 
\!\!\!\mbox{(P1-F-eq)}\quad{\underset{\mv{R}_x}{\rm max}}\quad & 0 \\ 
{\rm{s.t.}} \quad & {\rm{tr}}\left(\mv{R}_x\right) \leq P \\ 
&\begin{bmatrix}
	{\rm{tr}}(\mv{M}_2\mv{R}_x)\!-\!\xi& {\rm{tr}}(\mv{M}_3\mv{R}_x)\\
	{\rm{tr}}(\mv{M}_3^H\mv{R}_x)& {\rm{tr}}(\mv{M}_1\mv{R}_x)
\end{bmatrix} \!\succeq\! \mv{0}.
\end{align}
Note that Problem (P1-F-eq) is a semi-definite program (SDP), for which the optimal solution can be obtained via the interior-point method or software such as CVX \cite{bibConvexOpt}. It is worth noting that the feasibility of (P1) indicates the reachability of a certain PCRB level in sensing. In the following, we focus on the case where Problem (P1) has been verified to be feasible.

\subsection{Optimal Solution to Problem (P1)}\label{OP}
Note that the key difficulties in solving Problem (P1) lie in the secrecy rate expression in the objective function which involves logarithm functions of complicated fractional quadratic functions (i.e., the SINRs), as well as the complex PCRB constraint in (\ref{P1c2}). In the following, we overcome these challenges via a \emph{two-stage method}. \emph{Firstly}, we will consider a fixed threshold for the eavesdropping SINR at all possible target locations, which simplifies the fractional functions in the original objective function. Based on this, we will derive the optimal beamforming design by exploiting the problem structure. \emph{Secondly}, we will obtain the optimal SINR threshold via one-dimensional search, based on which the globally optimal beamforming design will be obtained.

Specifically, we introduce an auxiliary variable $\gamma$ to characterize the eavesdropping SINR threshold at each possible target location. It can be shown that there always exists a $\gamma>0$ at all possible eavesdropper locations such that Problem (P1.1) below has the same optimal solution to Problem (P1):
\begin{align} \label{con: sub1}
\!\!\!\!({\mbox{P1.1}}){\underset{\mv{w},\{\mv{v}_j\}_{j=1}^J}{\rm max}}\ & {{{\frac{|{\mv{h}^{H}\mv{w}}|^{2}}{\sum_{j=1}^J|\mv{h}^{H}\mv{v}_j|^{2} + \sigma^{2}}} }} \\ 
{\rm{s.t.}}\ &  \frac{|\mv{a}(\theta_k)^{H}\mv{w}|^{2}}{\sum_{j=1}^J|\mv{a}(\theta_k)^{H}\mv{v}_j|^{2}\!+\! \frac{\sigma_{\mathrm{E}}^2r^2}{\beta_0}}\leq \gamma,\forall k\!\in \!\mathcal{K}\!\\ 
& {\left\| \mv{w} \right\|^{2} + \sum_{j=1}^J\|\mv{v}_j\|^{2} \leq P} \\ 
& \mv{w}^H\mv{M}_2\mv{w}
\!+\!\sum_{j = 1}^J\mv{v}_j^H\mv{M}_2\mv{v}_j\nonumber\\
&\!-\!\frac{|\mv{w}^H\mv{M}_3\mv{w}\!+\!\sum_{j = 1}^J\mv{v}_j^H\mv{M}_3\mv{v}_j|^2}{\mv{w}^H\mv{M}_1\mv{w}\!+\!\sum_{j = 1}^J\mv{v}_j^H\mv{M}_1\mv{v}_j} \!\geq\xi.
\end{align}
Denote $f(\gamma)$ as the optimal value of Problem (P1.1) with given $\gamma>0$. Then, the following problem can be shown to have the same optimal value as Problem (P1) \cite{bib3}:
\begin{align} \label{con: sub2}
({\mbox{P1.2}}) \quad{\underset{\gamma>0}{{\rm{max}}}} \quad \log_2\left(\frac{1+f(\gamma)}{1+\gamma}\right).
\end{align}
Therefore, the optimal solution to Problem (P1) can be obtained via one-dimensional search of $\gamma>0$ in (P1.2) based on the values of $f(\gamma)$. Thus, our remaining task is to obtain $f(\gamma)$ by solving Problem (P1.1).

Motivated by the quadratic functions in (P1.1), we define $\mv{H}\overset{\Delta}{=}\mv{h}\mv{h}^{H}$, $\mv{W}\overset{\Delta}{=}\mv{w}\mv{w}^{H}$, $\mv{V}\overset{\Delta}{=}\sum_{j=1}^J\mv{v}_j\mv{v}_j^H$, and $\mv{A}_{k} \overset{\Delta}{=}\mv{a}(\theta_k)\mv{a}^H(\theta_k),\forall k$. Then, (P1.1) can be equivalently expressed as the following SDP with an additional constraint of $\mathrm{rank}(\mv{W})=1$:
\begin{align}
\mbox{(P1.1R)}&\nonumber\\
{\underset{\mv{W},\mv{V}}{{\rm{max}}}} \quad  & { \frac{{\rm{tr}}\left( {\mv{H}\mv{W}} \right)}{{\rm{tr}}\left( \mv{H}\mv{V} \right) + \sigma^{2}}}  \\ 
{\rm{s.t.}} \quad & {\rm tr}\left( {\mv{A}_{k}\mv{W}} \right)\! \leq\! \gamma\left( {\rm{tr}}\left({\mv{A}_k\mv{V}}\right)\! +\! \frac{\sigma_{\mathrm{E}}^2r^2}{\beta_0} \right),\forall k\in \mathcal{K}\label{P1.1Rc1}\\ 
& {\rm{tr}}\left( \mv{W} \right) + {\rm{tr}}\left( \mv{V} \right) \leq P   \label{P1.1Rc2}\\
&{\rm{tr}}(\mv{M}_2(\mv{W}+\mv{V}))-\frac{|{\rm{tr}}(\mv{M}_3(\mv{W}+\mv{V})|^2}{{\rm{tr}}(\mv{M}_1(\mv{W}+\mv{V}))}\geq \xi\label{P1.1Rc3}\\ 
& \mv{W} \succeq \mv{0}\label{P1.1Rc4}\\
&\mv{V} \succeq \mv{0}.\label{P1.1Rc5}
\end{align}	
Following the similar Schur complement procedures as in Section \ref{sec_feas} with $\mv{R}_x$ replaced by $\mv{W}+\mv{V}$, (P1.1R) can be shown to be equivalent to the following problem:
\begin{align}
\mbox{(P2.1R)}&\nonumber\\
{\underset{\mv{W},\mv{V}}{{\rm{max}}}}\quad  & { \frac{{\rm{tr}}\left( {\mv{H}\mv{W}} \right)}{{\rm{tr}}\left( \mv{H}\mv{V} \right) + \sigma^{2}}}  \\ 
{\rm{s.t.}} \quad & (\ref{P1.1Rc1}), (\ref{P1.1Rc2}), (\ref{P1.1Rc4}), (\ref{P1.1Rc5})\\
&\begin{bmatrix}
	{\rm{tr}}(\mv{M}_2(\mv{W}\!\!+\!\!\mv{V}))\!-\!\xi& {\rm{tr}}(\mv{M}_3(\mv{W}\!\!+\!\!\mv{V}))\\
	{\rm{tr}}(\mv{M}_3^H(\mv{W}\!\!+\!\!\mv{V}))& {\rm{tr}}(\mv{M}_1(\mv{W}\!\!+\!\!\mv{V}))
\end{bmatrix} \!\succeq\! \mv{0}.
\end{align}

Note that the objective function of (P2.1R) is still non-concave. To address this issue, we leverage the Charnes-Cooper transformation \cite{CharnesCooperT} to transform (P2.1R) into an equivalent convex problem as:
\begin{align}
\mbox{(P3.1R)}&\nonumber\\
{\underset{\mv{W},\mv{V},t}{{\rm{max}}}} \  & {\rm{tr}}\left( {\mv{H}\mv{W}} \right)  \\ 
{\rm{s.t.}} \ &{\rm tr}\left( {\mv{A}_{k}\mv{W}} \right)\! \leq\! \gamma\left( {\rm{tr}}\left({\mv{A}_k\mv{V}}\right)\! +\! \frac{t\sigma_{\mathrm{E}}^2r^2}{\beta_0} \right), \forall k\in \mathcal{K}\label{P3.1c1}\\ 
&{\rm{tr}}\left( \mv{H}\mv{V} \right)+t\sigma^2 = 1 \label{P3.1c2}\\ 
& {\rm{tr}}\left( \mv{W} \right) + {\rm{tr}}\left( \mv{V} \right) \leq tP  \label{P3.1c3}\\  \label{P3.1cF}
&\begin{bmatrix}
	{\rm{tr}}(\mv{M}_2(\mv{W}\!\!+\!\!\mv{V}))\!-\!t\xi& {\rm{tr}}(\mv{M}_3(\mv{W}\!\!+\!\!\mv{V}))\\
	{\rm{tr}}(\mv{M}_3^H(\mv{W}\!\!+\!\!\mv{V}))& {\rm{tr}}(\mv{M}_1(\mv{W}\!\!+\!\!\mv{V}))
\end{bmatrix} \!\succeq\! \mv{0}\!\!\!\\ 
&(\ref{P1.1Rc4}), (\ref{P1.1Rc5})\\
&t>0.\label{P3.1c5}
\end{align}

Note that Problem (P3.1R) is a convex optimization problem, for which the optimal solution can be obtained via the interior-point method or CVX \cite{bibConvexOpt}. Moreover, the Slater's condition is satisfied, thus the duality gap equals to zero \cite{bibConvexOpt}. In the following, we leverage the Lagrange duality method to unveil useful properties of the optimal solution to Problem (P3.1R), in order to reveal its relationship with the optimal solution to Problem (P1).

Let $\{\beta_k\geq 0\}_{k=1}^K$, $\lambda\geq 0$, $\rho\geq 0$, and $\mv{Z}\succeq \mv{0}$ denote the dual variables associated with constraints in (\ref{P3.1c1}), (\ref{P3.1c2}), (\ref{P3.1c3}) and (\ref{P3.1cF}), respectively, where $\mv{Z}= \begin{bmatrix}
z_{11} & z_{12}\\ 
z_{12}^{*} & z_{22}
\end{bmatrix}$. 
The Lagrangian of Problem (P3.1R) can be expressed as
\begin{align}\label{Lagrangian}
\mathcal{L}(\mv{W},\mv{V},t,\{\beta_k\},\lambda,\rho,\mv{Z}) = {\rm{tr}}(\mv{B}_1\mv{W})+{\rm{tr}}(\mv{B}_2\mv{V})+\omega t+\lambda,
\end{align}
where 
\begin{align}
\mv{B}_1&= \mv{H}-\sum_{k=1}^K\beta_k\mv{A}_k-\rho\mv{I}_{N_t}+\mv{B}_3, \\	\mv{B}_2&=-\lambda\mv{H}+\gamma\sum_{k=1}^K\beta_k\mv{A}_k-\rho\mv{I}_{N_t}+\mv{B}_3,\\
\mv{B}_3&=z_{11}\mv{M}_2+2{\mathfrak{Re}}\left\{z_{12}^{*}\mv{M}_3\right\}+z_{22}\mv{M}_1,\\
\omega&=-\lambda\sigma^2+\gamma\frac{\sigma_{\mathrm{E}}^2 r^2}{\beta_0}\sum_{k=1}^K\beta_k+\rho P-z_{11}\xi.
\end{align}
Let $\lambda^\star$, $\{\beta_k^\star\}_{k=1}^K$, $\rho^\star$, and $\mv{Z}^\star$ denote the optimal dual variables to (P3.1R). Define $\mv{D}^\star=-\lambda^\star\mv{H}-\sum_{k=1}^K\beta_k^\star\mv{A}_k-\rho^\star\mv{I}_{N_t}+z^\star_{11}\mv{M}_2+2{\mathfrak{Re}}\left\{z_{12}^{*^\star}\mv{M}_3\right\}+z_{22}^\star\mv{M}_1$, and $l=\mathrm{rank}(\mv{D}^\star)$. The orthogonal basis of the null space of $\mv{D}^{\star}$ can be represented as $\mv{U}\in \mathbb{C}^{N_t\times(N_t-l)}$, where $\mv{u}_{n}$ denotes the $n$-th column of $\mv{U}$. Then, we have the following proposition.
\begin{proposition}\label{prop_rank}
For Problem (P3.1R), the optimal $\mv{W}^\star$ can be expressed as follows with $a_n\geq 0,\forall n$, $b>0$, and $\mv{r}$ satisfying $\mv{r}^H\mv{U} = \mv{0}$:
\begin{align}
	\mv{W}^\star = \sum_{n=1}^{N_t-l}a_n\mv{u}_n\mv{u}_n^H+b\mv{r}\mv{r}^H.
\end{align}
If ${\rm{rank}}(\mv{W}^\star) >1$, the following set of solution with a rank-one solution of $\mv{W}$ can be constructed which achieves the same optimal value of (P3.1R):
\begin{align}
	{\tilde{\mv{W}}}^\star = & b\mv{r}\mv{r}^H,\label{Ws}\\
	{\tilde{\mv{V}}}^\star = & \mv{V}^\star + \sum_{n=1}^{N_t-l}a_n\mv{u}_{n}\mv{u}_{n}^H,\label{Vs}\\
	{\tilde{t}^\star} = & t^\star.\label{ts}
\end{align}
\end{proposition}
\begin{IEEEproof}
Please refer to Appendix \ref{proof_prop_rank}.
\end{IEEEproof}

The results in Proposition \ref{prop_rank} indicate that the relaxation from Problem (P1.1) to Problem (P3.1R) as well as Problem (P1.1R) is tight. Therefore, we can first obtain the optimal solution to Problem (P3.1R) denoted by $(\mv{W}^\star,\mv{V}^\star,t^\star)$ via interior-point method or CVX. If $\mathrm{rank}(\mv{W}^\star)=1$, the optimal solution to Problem (P1.1) can be obtained via the eigenvalue decompositions (EVDs) of $\mv{W}^\star$ and $\mv{V}^\star$; otherwise, $(\tilde{\mv{W}}^\star,\tilde{\mv{V}}^\star)$ can be constructed based on (\ref{Ws}) and (\ref{Vs}), and the optimal solution to Problem (P1.1) can be obtained via the EVDs of $\tilde{\mv{W}}^\star$ and $\tilde{\mv{V}}^\star$. Then, by obtaining the optimal $\gamma$ via solving Problem (P1.2) using one-dimensional search, the optimal solution to Problem (P1) can be obtained.

\subsection{Suboptimal Solutions to Problem (P1)}
In this subsection, we propose two suboptimal solutions to Problem (P1), which are of lower complexity compared to the optimal solution.
\subsubsection{Suboptimal Solution I}
In this suboptimal solution, the optimal beamforming vectors for the information beam and the AN beams lie in the null spaces of the eavesdropping target's possible channels and the user channel, respectively. Note that this achieves zero ``interference'' between the user and the eavesdropper; while on the other hand, the information beam cannot contribute to the echo signals for sensing, which makes this solution suboptimal in general.

First, we express the singular value decomposition (SVD) of the collection of all possible BS-eavesdropper channels $\mv{A}=[\mv{a}(\theta_1),...,\mv{a}(\theta_K)]^H$ as $\mv{A}= \mv{U}_A\mv{\Lambda}_A\left[\mv{J}_1\ \mv{J}_2\right]^H$, 
where $\mv{U}_A\in\mathbb{C}^{K\times Q}$ with $Q=\mathrm{rank}(\mv{A})$ and $\left[\mv{J}_1~\mv{J}_2\right]^H\in\mathbb{C}^{N_t\times N_t}$ are unitary matrices; $\mv{\Lambda}_A$ is a $Q\times N_t$ rectangular diagonal matrix; $\mv{J}_1\in \mathbb{C}^{N_t\times Q}$ and $\mv{J}_2\in \mathbb{C}^{N_t\times (N_t-Q)}$ consist of the first $Q$ and the last $N_t-Q$ right singular vectors of $\mv{A}$, respectively. 
In addition, $\mv{J}_2$ forms an orthogonal basis for the null space of $\mv{A}$. Thus, to guarantee  $\mv{A}\mv{w}=\mv{0}$, the information beam $\mv{w}$ should satisfy 
\begin{align}
\mv{w}=\sqrt{{P}_w}\mv{J}_2\tilde{\mv{w}},
\end{align} 
where ${P}_w=\|\mv{w}\|^2$ denotes the transmit power allocated to the information beam, and $\tilde{\mv{w}}\in\mathbb{C}^{\left(N_t-Q\right)\times 1}$ with $\|\tilde{\mv{w}}\|=1$ can be arbitrarily designed. To maximize the power of the information beam at the user, we align $\tilde{\mv{w}}$ with $\mv{J}_2^H\mv{h}$ as $\tilde{\mv{w}}^\star=\frac{\mv{J}_2^H\mv{h}}{\|\mv{J}_2^H \mv{h}\|}$.

Next, to guarantee $\mv{h}^H\mv{v}_j=0,\forall j$, we define $\mv{T}_s=\mv{I}_{N_t}-\frac{\mv{h}\mv{h}^H}{\|\mv{h}\|^2}$, which can be expressed as $\mv{T}_s = \tilde{\mv{X}}\tilde{\mv{X}}^H$, with $\tilde{\mv{X}}\in \mathbb{C}^{N_t\times (N_t-1)}$ forming an orthogonal basis for the null space of $\mv{h}^H$. Thus, $\mv{v}_j$ must satisfy
\begin{align}
\mv{v}_j=\tilde{\mv{X}}\tilde{\mv{v}}_j,~j = 1,..., J,
\end{align} 
where $\tilde{\mv{v}}_j\in\mathbb{C}^{(N_t-1)\times 1}$ can be arbitrarily designed. Under this solution structure, the achievable secrecy rate is simplified as
\begin{align}\label{r0A}
r_0^{\mathrm{I}}={\rm{log}}_2\left(1+\frac{{P}_w\|{\mv{J}_2}^H\mv{h}\|^2}{\sigma^2}\right),
\end{align}
which is maximized when $P_w$ is maximized. On the other hand, the PCRB constraint in (\ref{P1c2}) reduces to
\begin{align}\label{PCRB_subI}
P_wM_2+\sum_{j = 1}^J\tilde{\mv{v}}_j^H\tilde{\mv{M}}_2\tilde{\mv{v}}_j
\!-\!\frac{|P_wM_3\!+\!\sum_{j=1}^J\tilde{\mv{v}}_j^H\tilde{\mv{M}}_3\tilde{\mv{v}}_j|^2}{P_wM_1+\sum_{j=1}^J\tilde{\mv{v}}_j^H\tilde{\mv{M}}_1\tilde{\mv{v}}_j}\leq \xi,
\end{align}
where $M_2 =\tilde{\mv{w}}^{\star^H}\mv{F}_2^H \mv{M}_2\mv{F}_2\tilde{\mv{w}}^{\star}$, $M_3=\tilde{\mv{w}}^{\star^H}\mv{F}_2^H \mv{M}_3\mv{F}_2\tilde{\mv{w}}^{\star}$, $M_1 =\tilde{\mv{w}}^{\star^H}\mv{F}_2^H \mv{M}_1\mv{F}_2\tilde{\mv{w}}^{\star}$, $\tilde{\mv{M}}_2=\tilde{\mv{X}}^H\mv{M}_2\tilde{\mv{X}}$, $\tilde{\mv{M}}_3=\tilde{\mv{X}}^H\mv{M}_3\tilde{\mv{X}}$ and $\tilde{\mv{M}}_1=\tilde{\mv{X}}^H\mv{M}_1\tilde{\mv{X}}$. 

By defining $\tilde{\mv{V}}=\sum_{j=1}^J\tilde{\mv{v}}_j\tilde{\mv{v}}_j^H$ and applying the Schur complement technique on (\ref{PCRB_subI}), (P1) is reduced to the following joint optimization problem of the power allocated to the information beam, $P_w$, and the AN beam design:
\begin{align}
\mbox{(P1-Sub-I)}&\nonumber\\
 \underset{P_w\geq 0,\tilde{\mv{V}}\succeq \mv{0}}{\max}\ &P_w\\ \label{AP1Rcons1}
{\rm{s.t.}}\ &\begin{bmatrix}
	P_wM_2\!+\!{\rm{tr}}(\tilde{\mv{M}}_2\tilde{\mv{V}})\!\!-\xi&\!\! {P_wM_3\!\!+\!\!\rm{tr}}\!(\tilde{\mv{M}}_3\tilde{\mv{V}}\!)\\
	P_wM_3^{*}\!\!+\!\!{\rm{tr}}(\tilde{\mv{M}}_3^H\tilde{\mv{V}}\!)\!\!&\!\! P_wM_1\!\!+\!\!{\rm{tr}}\!(\!\tilde{\mv{M}}_1\tilde{\mv{V}}\!)
\end{bmatrix}\succeq \mv{0}\\ 
& P_w+{\rm{tr}}(\tilde{\mv{V}})\leq P.
\end{align}
The above problem is convex, for which the optimal solution denoted by $(P_w^\star,\tilde{\mv{V}}^\star)$ can be obtained via CVX. The optimal solution of $\tilde{\mv{v}}_j$'s to (P1-Sub-I) denoted by $\tilde{\mv{v}}^\star_j$ can be obtained via the EVD of $\tilde{\mv{V}}^\star$. Consequently, the AN beamforming vectors are designed as 
\begin{align} \label{con: benchmark1_v}
\mv{v}_j^\star=\tilde{\mv{X}}\tilde{\mv{v}}_j^\star,~~j = 1,...,J.
\end{align}
The information beamforming vector is designed as
\begin{align}
\mv{w}^\star = \sqrt{P_w^\star}\mv{J}_2\tilde{\mv{w}}^\star.
\end{align}

\subsubsection{Suboptimal Solution II}
In this suboptimal solution, we directly align the information beam with the user channel $\mv{h}$, i.e., $\mv{w}=\frac{\sqrt{\hat{P}_w}\mv{h}}{\|\mv{h}\|}$ with $\hat{P}_w$ denoting the power allocated to the information beam, and design the AN beams such that they generate zero interference to the user with $\mv{v}_j^H\mv{h} = 0,j=1,...,J$ in a similar manner to that in suboptimal solution I as
\begin{align}\hat{\mv{v}}_j=\sqrt{P-\hat{P}_w}\frac{\mv{v}_j^\star}{\sqrt{{\sum_{j=1}^J\|\mv{v}_j^\star\|^2}}},~j = 1,..., J.
\end{align}
Note that compared to suboptimal solution I, although suboptimal solution II also aims to minimize the interference of the AN beams to the user, it aims to maximize the user's desired signal power instead of minimizing the information leakage to the eavesdropper. Under this structure, the achievable secrecy rate is given by
\begin{align}
r_0^{\mathrm{II}} = &\underset{k\in \mathcal{K}}{\rm{min}}~{\rm{log}}_2\left(1+\frac{\hat{P}_w\|\mv{h}\|^2}{\sigma^2}\right)\!\\ 
&-\!{\rm{log}}_2\left(1+\frac{\hat{P}_w|\mv{h}^H\mv{a}(\theta_k)|^2}{\|\mv{h}\|^2\left(\!\frac{(P-\hat{P}_w)\sum_{j= 1}^J|\mv{a}(\theta_k)^H\mv{v}_j^\star|^2}{\sum_{j=1}^J\|\mv{v}_j^\star\|^2}+\frac{r^2\sigma_{\mathrm{E}}^2}{\beta_0}\right)}\right).\nonumber
\end{align}
The PCRB constraint can be expressed as
\begin{align}\label{PCRB_subII}
&\hat{P}_w\frac{\mv{h}^H\mv{M}_2\mv{h}}{\|\mv{h}\|^2}+(P-\hat{P}_w)\frac{\sum_{j=1}^J\mv{v}_j^{*H}\mv{M}_2\mv{v}_j^{*}}{\sum_{j=1}^J\|\mv{v}_j^\star\|^2}\nonumber\\ -
&\frac{\left|\hat{P}_w\frac{\mv{h}^H\mv{M}_3\mv{h}}{\|\mv{h}\|^2}+(P-\hat{P}_w)\frac{\sum_{j=1}^J\mv{v}_j^{*H}\mv{M}_3\mv{v}_j^{*}}{\sum_{j=1}^J\|\mv{v}_j^\star\|^2}\right|^2}{\hat{P}_w\frac{\mv{h}^H\mv{M}_1\mv{h}}{\|\mv{h}\|^2}+(P-\hat{P}_w)\frac{\sum_{j=1}^J\mv{v}_j^{*H}\mv{M}_1\mv{v}_j^{*}}{\sum_{j=1}^J\|\mv{v}_j^\star\|^2}}\leq \xi.
\end{align}
Thus, (P1) reduces to the following problem under the proposed structure:
\begin{align}
\mbox{(P1-Sub-II)}\quad {\underset{\hat{P}_w: (\ref{PCRB_subII})}{\rm{max}}}\quad & r_0^{\mathrm{II}}.
\end{align}
The optimal solution $\hat{P}_w^\star$ to Problem (P1-Sub-II) can be obtained via one-dimensional search over $[0,P]$.

\subsection{Complexity Analysis}
In the following, we analyze the computational complexity for our proposed solutions. To obtain the optimal solution, Problem (P3.1R) should be solved first, in which there are two matrix variables with size $N_t\times N_t$ and one scalar variable, thus the total number of optimization variables is $m=2N_t^2+1$. There are two positive-semidefinite constraints with size $N_t\times N_t$, one positive-semidefinite constraint with size $2\times 2$, and $K+3$ linear constraints, thus the number of operations in the interior-point method is in the order of $n = {\rm{log}}\tilde{\epsilon}^{-1}{\sqrt{2N_t+K+7}}$, where $\tilde{\epsilon}$ denotes the accuracy requirement. Thus, the overall complexity to solve Problem (P3.1R) is calculated as $\mathcal{O}(n(m(2N_t^3+2^3+K+3)+m^2(2N_t^2+2^2+K+3)+m^3))$. By further taking into account the complexity in obtaining the optimal solution to Problem (P1.1) via EVD and the complexity in one-dimensional search over $\gamma$ with $N_o$ sample points, the complexity for obtaining the optimal solution to (P1) is $\mathcal{O}(N_{o}n(m(2N_t^3+K+11)+m^2(2N_t^2+K+7)+m^3)+2N_t^3)\propto\mathcal{O}(N_o{\rm{log}}\tilde{\epsilon}^{-1}N_t^{6.5})$. Similarly, the  complexity for obtaining suboptimal solution I can be shown to be $\mathcal{O}({\rm{log}}\tilde{\epsilon}^{-1}{\sqrt{N_t+1}}((N_t-1)((N_t-1)^3+2)+(N_t-1)^2((N_t-1)^2+2)+(N_t-1)^3)+KN_t^2+N_t^3+(N_t-1)^3)\propto \mathcal{O}({\rm{log}}\tilde{\epsilon}^{-1}N_t^{4.5})$. For suboptimal solution II, with $N_s$ denoting the sample points in the one-dimensional search of $\hat{P}_w$, the complexity is $\mathcal{O}({\rm{log}}\tilde{\epsilon}^{-1}{\sqrt{N_t+1}}((N_t-1)((N_t-1)^3+2)+(N_t-1)^2((N_t-1)^2+2)+(N_t-1)^3)+KN_t^2+N_t^3+(N_t-1)^3+N_s)$.

\section{Numerical Results}\label{SectionSimulation}
In this section, we provide numerical results to evaluate the performance of the proposed beamforming designs for secure ISAC. We set $N_t = 8$ transmit antennas and $N_r=10$ receive antennas for the BS. As illustrated in Fig. \ref{Figpattern}, we consider $K=4$ possible target (eavesdropper) locations, with $\theta_1=-1.22,\theta_2=-0.79,\theta_3=-0.44$, $\theta_4=0.87$ (in radian); $p_1=0.2,p_2=0.1,p_3=0.4$, and $p_4=0.3$; $\sigma_{\theta}^2=10^{-4}$. The transmit power is set as $P=20$ dBm. The path loss of the BS-target channel is set as $40$ dB. The lower bound of the target's RCS gain is set as $|\bar{\alpha}|=0.32$. The average receiver noise power is set as $\sigma_{\mathrm{R}}^2=\sigma^2_{\mathrm{E}} = \sigma^2 = -80~{\rm{dBm}}$. The BS-user channel is assumed to follow the Rayleigh fading model with $\mv{h}\sim\mathcal{CN}(\mv{0},\sigma_h^2\mv{I})$ where $\sigma_h^2=-80~{\rm{dB}}$. Note that this is a challenging scenario for secure communication where the user is located farther away from the BS with a statistically weaker channel compared to the eavesdropper.

\subsection{Performance of Secure ISAC with Optimal Beamforming}
\begin{figure}[t]
\centering
\includegraphics[width=9cm]{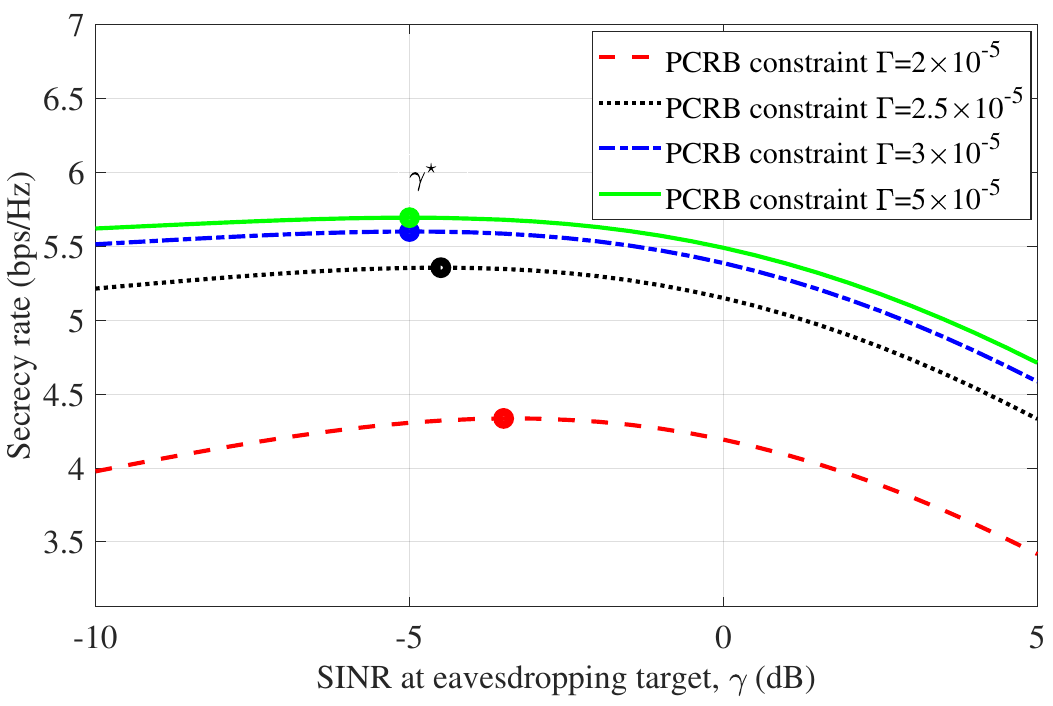}
\caption{Secrecy rate versus the SINR constraint at the eavesdropper, $\gamma$, under different PCRB constraints.}
\label{FigSRvsgamma}
\end{figure}

\begin{figure}[t]
\centering
\includegraphics[width=9cm]{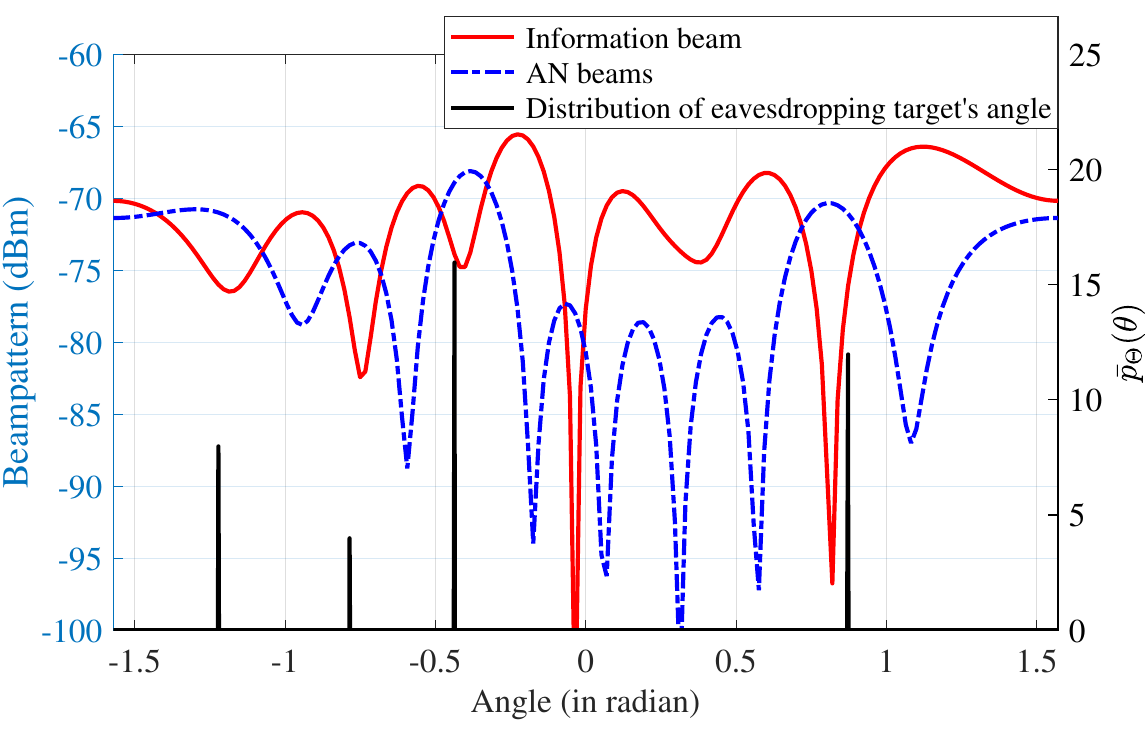}
\caption{Beampattern and target location distribution $\bar{p}_{\Theta}(\theta)$ over different angles.}
\label{Figpattern}
\end{figure}

To start with, we consider one realization of $\mv{h}$ and evaluate the performance of the optimal solution. In Fig. \ref{FigSRvsgamma}, we illustrate  ${\rm{log}}_2((1+f(\gamma))/(1+\gamma))$ versus the SINR constraint at the eavesdropper, $\gamma$, under various sensing accuracy threshold $\Gamma$. It is observed that there exists a unique optimal value of $\gamma$ for every $\Gamma$; moreover, the optimal value decreases as the PCRB constraint becomes less stringent, leading to higher secrecy rate. Then, we consider $\Gamma = 3\times 10^{-5}$ and illustrate in Fig. \ref{Figpattern} the beampattern over different angles at a distance with path loss $80$ dB. It is observed that the power of the information beam reaches local minimums at the possible target locations; while the power of the AN beams is high over the possible target locations. Moreover, the total power focused at each possible target location generally increases as the target's associated probability increases. This demonstrates the efficacy of the optimal solution in simultaneously avoiding information leakage and achieving probability-dependent power focusing.

\begin{figure}[t]
	\centering
	\includegraphics[width=8cm]{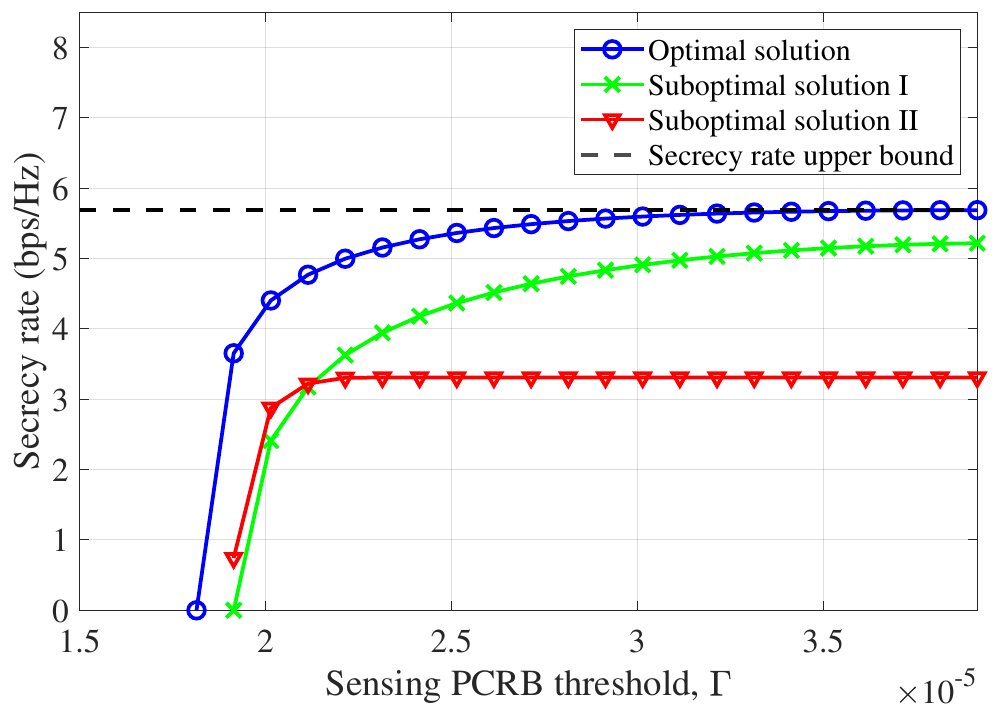}
	\caption{Secrecy rate versus sensing PCRB threshold $\Gamma$.}
	\label{FigCompare}
\end{figure}
\begin{figure}[t]
	\centering
\includegraphics[width=8cm]{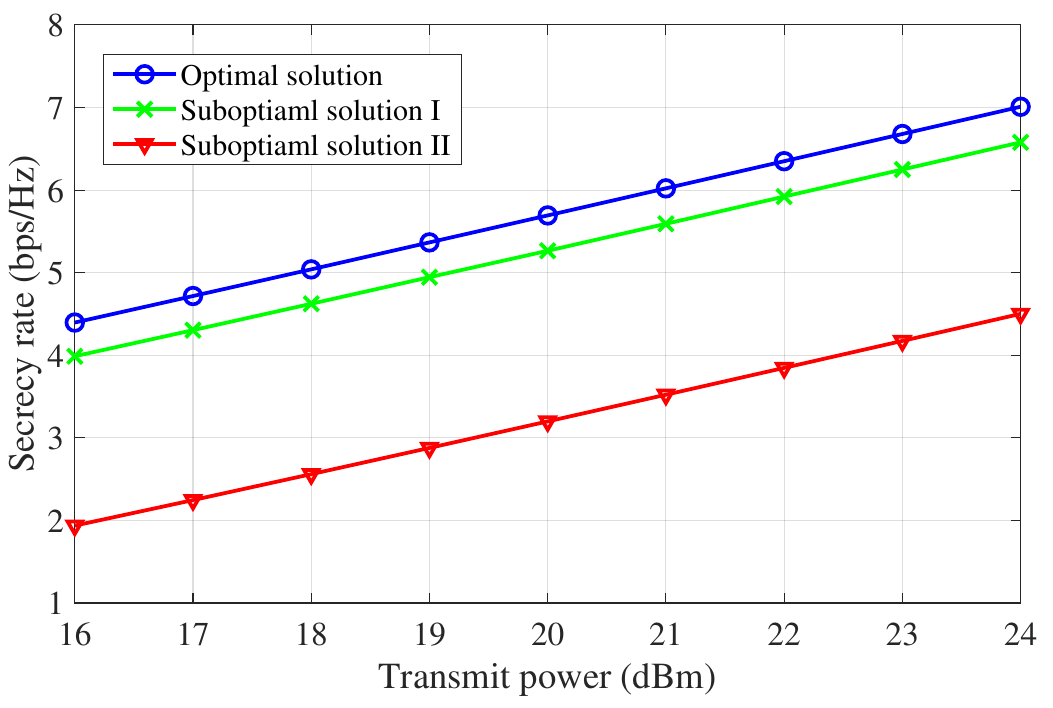}
\caption{Secrecy rate versus transmit power under $\Gamma=7\times 10^{-5}$.}
\label{FigTransmitPower}
\vspace{-6mm}\end{figure}
\subsection{Comparison Between Optimal and Suboptimal Solutions}
Next, we compare the performance and complexity between the optimal solution and the proposed two suboptimal solutions, where the results are averaged over $50$ independent user channel realizations. In Fig. \ref{FigCompare}, we show the secrecy rate versus the PCRB threshold $\Gamma$ of the proposed optimal and suboptimal solutions, as well as an upper bound of the secrecy rate  where no sensing function and PCRB constraint is considered. It is observed that the optimal solution achieves significantly improved performance compared with the two suboptimal solutions, and approaches the upper bound as the PCRB constraint becomes less stringent, which demonstrates the effectiveness of beamforming optimization. Moreover, suboptimal solution I generally outperforms suboptimal solution II due to the stringent requirement of achieving zero information leakage. It is worth noting that suboptimal solution II achieves better performance only when the PCRB constraint is very tight (i.e., when $\Gamma$ is very small). In this case, it is more desirable for the information beam to make non-zero contribution to the echo signal strength as in suboptimal solution II. Furthermore, for all three solutions, there exists a trade-off between the secrecy communication performance and the sensing accuracy.

\begin{figure}[t]
\centering
\includegraphics[width=8cm]{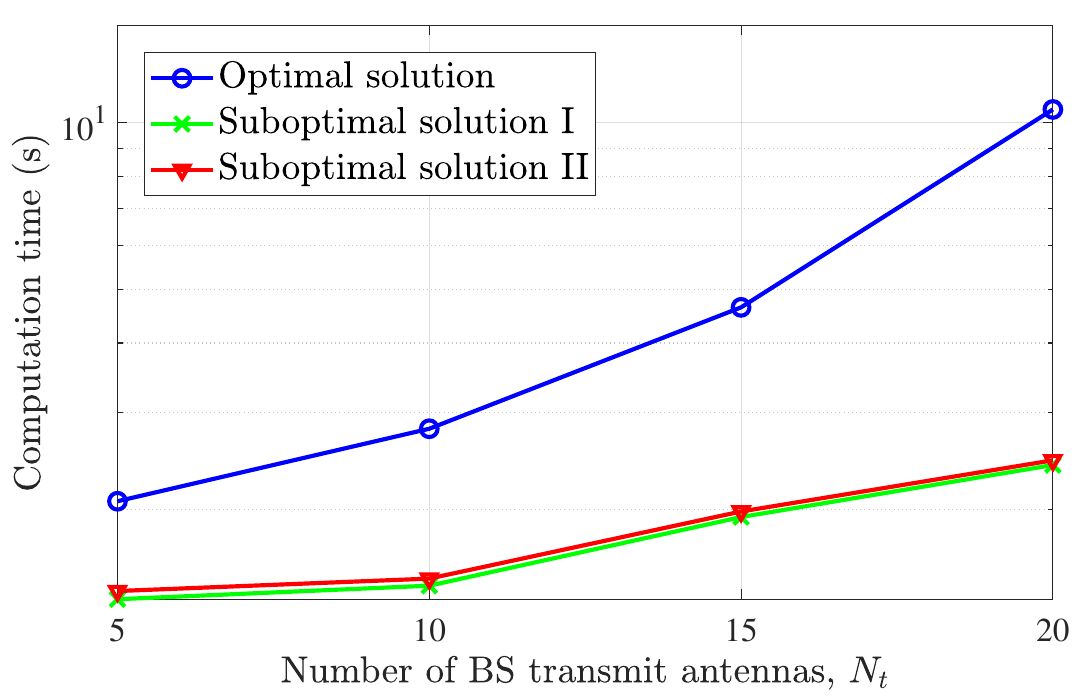}
\vspace{-3mm}
\caption{Computation time versus the number of BS transmit antennas.}
\label{FigComplexity}
\vspace{-6mm}\end{figure}

In Fig. \ref{FigTransmitPower}, we show the secrecy rate of the proposed solutions versus the transmit power $P$ with PCRB constraint $\Gamma = 7\times10^{-5}$. It is observed that the optimal solution achieves $1$ dB power gain over suboptimal solution I, and around $8$ dB power gain over suboptimal solution II. This further shows the efficacy of the optimal solution. Finally, we show in Fig. \ref{FigComplexity} the computation time of the proposed solutions obtained by MATLAB on a computer with an Intel Core i7 2.50-GHz CPU and 16 GB of memory. It is observed that compared to the suboptimal solutions, the optimal solution requires significantly higher computation time as the number of BS transmit antennas, $N_t$, becomes larger; moreover, suboptimal solution II requires slightly higher complexity compared to suboptimal solution I due to the extra one-dimensional search required. The results are consistent with our complexity analysis in Section V-D. 
\section{Conclusions}\label{SectionConclusion}
This paper studied the transmit beamforming optimization in a challenging secure ISAC scenario where the location parameter of the sensing target which also serves as a potential eavesdropper is unknown and random, whose PMF is available for exploitation. First, to characterize the sensing performance exploiting discrete PMF, we proposed a Gaussian mixture model based PDF to approximate the PMF, based on which we derived the PCRB for target sensing. To draw more insights, we further proposed a tight approximation of the PCRB in closed form, which implies a ``probability-dependent power focusing'' guideline for the transmit beamforming design. Then, under an AN-based beamforming structure, we formulated the joint optimization problem of the information beam and AN beams at the transmitter to maximize the worst-case secrecy rate among all possible target locations, under a constraint on the exact sensing PCRB. The formulated problem is non-convex and difficult to solve. By applying advanced optimization techniques such as Schur complement, Charnes-Cooper transformation, and SDR, we proposed a two-stage algorithm to obtain its optimal solution with polynomial-time complexity. Moreover, we proposed two suboptimal solutions with lower complexity by designing the information and/or AN beams in the null spaces of the possible eavesdropper channels and/or the user channel, respectively. Numerical results validated the efficacy of the proposed solutions in achieving a favorable trade-off between secrecy communication and sensing exploiting target location distribution.

\appendix
\subsection{Proof of Proposition \ref{prop_app}}\label{proof_prop_app}
First, we denote $\mv{S}(\theta_k)=\int_{-\infty}^{\infty}f_k(\theta)\|\dot{\mv{b}}(\theta)\|^2\mv{a}(\theta)\mv{a}(\theta)^H{\rm{d}}\theta$, which yields $\mv{Q}=\sum_{k=1}^{K}\mv{S}(\theta_k)$. Based on the definition of $\mv{a}(\theta)$ and $\mv{b}(\theta)$, $\mv{S}(\theta_k)$ can be further expressed as
\begin{align}\label{cons: S}
&\mv{S}(\theta_k)=\frac{\sum_{n=1}^{N_r}\pi^2(n-1)^2}{2}\int_{-\infty}^{\infty} f_k(\theta){{\rm{cos}}^2\theta}\\
&{\begin{bmatrix}
		1&\!\!\!\!e^{-j\pi\sin\theta}&\!\!\!\!... &\!\!\!\!e^{-j\pi(N_t-1)\sin\theta}\\ 
		e^{j\pi\sin\theta} &\!\!\!\! 1 & \!\!\!\!... &\!\!\!\! e^{-j\pi(N_t-2)\sin\theta}\\ 
		\vdots & \ddots&  &\vdots\\ 
		e^{j\pi(N_t-1)\sin\theta}&\!\!\!\!e^{j\pi(N_t-2)\sin\theta} &\!\!\!\!...&\!\!\!\! 1
\end{bmatrix}}{\rm{d}}\theta.\nonumber
\end{align}
Denote $\mu_1 = \frac{p_k}{\sqrt{2\pi}}(\sum_{n=1}^{N_r}\pi^2(n-1)^2/2)$, ${\mu}_2 = -\pi$, and $t\!=\!\theta\!-\!\theta_k$. Then, $[\mv{S}(\theta_k)]_{1,2}$ can be further simplified as:
\begin{align}
&[\mv{S}(\theta_k)]_{1,2}=\mu_1\frac{1}{\sigma_\theta}\int_{-\infty}^{\infty}{e}^{{-\frac{(\theta-\theta_k)^2}{2\sigma_{\theta}^2}}}{\rm{cos}}^2\theta e^{- j\pi{\rm{sin}}\theta} {\rm{d}}\theta\\
&=\frac{\mu_1}{2\sigma_\theta}\int_{-\infty}^{\infty}{e}^{{-\frac{t^2}{2\sigma_{\theta}^2}}}e^{-j\pi(\sin\theta_k\cos t+\cos\theta_k\sin t)}\nonumber\\
&\qquad\quad\times(\cos (2t)\cos (2\theta_k)-\sin (2t)\sin (2\theta_k)+1 ) dt\nonumber\\
&\overset{(a)}{\approx}\![\tilde{\mv{S}}(\theta_k)]_{1,2}\!=\!\frac{\mu_1}{2\sigma_\theta}\int_{-\infty}^{\infty}{e}^{{-\frac{t^2}{2\sigma_{\theta}^2}}}(\alpha_0\!+\!\alpha_1t\!+\!\alpha_2t^2\!+\!o(t^3))dt,\nonumber
\end{align}
where $\alpha_0 = e^{-j\pi{\rm{sin}}(\theta_k)}({\rm{cos}}(2\theta_k)+1)$; $\alpha_1 = -2{\rm{sin}}(2\theta_k)+j\pi(1+{\rm{cos}}(2\theta_k)){\rm{sin}}(\theta_k)$; and $\alpha_2 = -2{\rm{cos}}(2\theta_k)+\frac{1}{2}\pi(j{\rm{sin}}(\theta_k)+{\rm{cos}}(\theta_k))({\rm{cos}}(\theta_k)+1)+2j\pi{\rm{sin}}(2\theta_k){\rm{cos}}(\theta_k)$. Note that $(a)$ is derived by taking the Maclaurin series of ${\rm{cos}}(2t)$, ${\rm{sin}}(2t)$, $e^{-j\pi{\rm{sin}}(\theta_k){\rm{cos}}t}$ and $e^{-j\pi{\rm{cos}}(\theta_k){\rm{sin}}t}$ and noting $\sigma^2_\theta$ is a small value. Notice that ${e}^{{-\frac{t^2}{2\sigma_{\theta}^2}}}t$ and ${e}^{{-\frac{t^2}{2\sigma_{\theta}^2}}}t^3$ are odd functions. Moreover, 
$\int_{-\infty}^{\infty}{e}^{{-\frac{t^2}{2\sigma_{\theta}^2}}}{\rm{d}}t\! =\! {{\sqrt{2\pi}}}\sigma_\theta $, and $\int_{-\infty}^{\infty}{e}^{{-\frac{t^2}{2\sigma_{\theta}^2}}}t^2{\rm{d}}t\!=\!\sqrt{2\pi}\sigma_\theta^3$. Thus, we have
$[\tilde{\mv{S}}(\theta_k)]_{1,2}=\mu_1({\rm{cos}}(2\theta_k)+1){\sqrt{\frac{\pi}{2}}}e^{-j\pi{\rm{sin}}(\theta_k)}$. 
Similarly, other entries in $\mv{S}(\theta_k)$ can be approximated in the same manner, which yields
\begin{align}\label{cons: S_value}
\tilde{\mv{S}}(\theta_k)= \mu_1\sqrt{\frac{\pi}{2}}({\rm{cos}}(2\theta_k)+1)\mv{a}(\theta_k)\mv{a}^H(\theta_k).
\end{align}

Secondly, due to the small value of $\sigma_\theta^2$, the non-zero values of $\frac{\partial \bar{p}_\Theta(\theta)}{\partial \theta}$ will only occur in the close vicinity of $\theta_k$'s. Thus, we have $\int_{-\infty}^{\infty}\frac{\Big(\frac{\partial  \bar{p}_\Theta(\theta)}{\partial \theta}\Big)^2}{\bar{p}_\Theta(\theta)}d\theta\approx \frac{1}{\sigma_\theta^2}$, i.e., $\epsilon\approx 0$. Based on this and (\ref{cons: S_value}), Proposition \ref{prop_app} is proved.

\subsection{Proof of Proposition \ref{prop_rank}}\label{proof_prop_rank}
First, we analyze the optimal values of $\lambda$ and $\rho$ in the following lemma to analyze their associated constraints.
\begin{lemma}
	The optimal dual variables to Problem (P3.1R) satisfy $\lambda^\star>0$ and $\rho^\star>0$.
\end{lemma}
\begin{IEEEproof}
To ensure that the Lagrangian in (\ref{Lagrangian}) is bounded so that the dual function exists, we should have $\mv{B}_1 \preceq\mv{0}$, $\mv{B}_2 \preceq\mv{0}$, and $\omega \leq0$. The dual problem of (P3.1R) can be expressed as
\begin{align}
	\mbox{(P3.1R-dual)}\underset{\{\beta_k\geq 0\}_{k=1}^K,\atop \lambda\geq 0, \rho\geq 0, \mv{Z}\succeq\mv{0}}{\min} \quad & \lambda\\
	\rm{s.t.} \qquad & \mv{B}_1\preceq \mv{0},\ \mv{B}_2\preceq \mv{0},\ \omega \leq 0.
\end{align}
Since the duality gap is zero, $\lambda^{\star}$ equals to the optimal value of (P3.1R). Thus, we have $\lambda^{\star}>0$. 

Next, under complimentary slackness, we have ${\mathrm{tr}}\left(\mv{Z}^\star\begin{bmatrix}
	{\rm{tr}}(\mv{M}_2(\mv{W}^\star\!\!+\!\!\mv{V}^\star))\!-\!t\xi& {\rm{tr}}(\mv{M}_3(\mv{W}^\star\!\!+\!\!\mv{V}^\star))\\
	{\rm{tr}}(\mv{M}_3^H(\mv{W}^\star\!\!+\!\!\mv{V}^\star))& {\rm{tr}}(\mv{M}_1(\mv{W}^\star\!\!+\!\!\mv{V}^\star))
\end{bmatrix}\right)=0$. This leads to ${\mathrm{det}}\left(\mv{Z}^\star\right) = 0$ and $z_{11}^\star z_{22}^\star - \left|z_{12}^\star\right|^2 = 0$. Since $\mv{Z}^\star\!\succeq\! \mv{0}$, $z_{11}^\star\!\geq\! 0$. Consequently, $z_{22}^\star\!\geq\! 0$, and ${\mathrm{det}}\left(\begin{bmatrix}\!
	z_{11}^\star\|\dot{\mv{b}}(\theta)\|+ z_{22}^\star N_r\!\!&\!\!z_{12}^\star N_r\!\! \\ 
	z_{12}^{*^\star} N_r\!\!&  \!\!z_{11}^\star N_r
\end{bmatrix}\right)\geq 0$. Thus, we have $\mv{B}_3^\star\succeq\mv{0}$ since $\mv{a}(\theta)^H \dot{\mv{a}}(\theta)=0$ and $\mv{B}_3=\int_{-\infty}^{\infty}\!\!\begin{bmatrix}\!
	\mv{a}(\theta)\dot{\mv{a}}(\theta)
\end{bmatrix}\begin{bmatrix}
	z_{11}\|\dot{\mv{b}}(\theta)\|+ z_{22}N_r\!\!&\!\!z_{12} N_r\!\! \\ 
	z_{12}^{*} N_r\!\!&  \!\!z_{11}N_r
\end{bmatrix}\!\!\begin{bmatrix}
	\mv{a}^H(\theta) \\ \dot{\mv{a}}^H(\theta)
\end{bmatrix}\!\!p_\Theta(\theta)d\theta$. 

Then, under Karush-Kuhn-Tucker (KKT) conditions, we have $\partial{\mathcal{L}(\mv{W},\mv{V},t,\{\beta_k\},\lambda,\rho,\mv{Z})}/\partial{t} =0$ at the optimal solution, which leads to $\sum_{k=1}^K\beta_k^{\star} \!= \!\frac{\beta_0}{\gamma\sigma_{\mathrm{E}}^2r^2}\!\!\left(\lambda^{\star}\sigma^2+z_{11}^{\star}\xi \!-\! \rho^{\star}P\right)$. If $\rho^{\star} = 0$, we have  $\sum_{k=1}^{K}\beta_k^{\star}>0$, and $\mv{B}_2^\star = -\lambda^{\star}\mv{H} + \gamma\sum_{k=1}^K\beta^\star\mv{A}_k +\mv{B}_3^\star$. Since $\mv{B}_2\preceq\mv{0}$ and $\mv{B}_3\succeq\mv{0}$, it can be observed $-\lambda^{\star}\mv{H} + \gamma\sum_{k=1}^K\beta^{\star}\mv{A}_k\preceq \mv{0}$. Since $\sum_{k=1}^{K}\beta_k^{\star}\mv{A}_k\succeq\mv{0}$ and $\lambda^{\star}>0$, to guarantee $\mv{B}_2\preceq\mv{0}$, any $\mv{x}\in \mathbb{C}^{N_t\times1}$ lying in the null space of $\mv{H}$ must also lie in that of $\sum_{k=1}^K\mv{A}_k$. However, since $\mv{h}$ and $\mv{a}(\theta_k)$'s are linearly independent, this cannot be true. Thus, $\rho^\star>0$ holds. 
\end{IEEEproof}	

With $\lambda^\star>0$ and $\rho^\star>0$, under KKT conditions, we have 
\begin{align}
	\mv{B}_1^{\star}\mv{W}^{\star} = \mv{0},\\
	\mv{B}_2^{\star}\mv{V}^{\star} = \mv{0},\\
	\mv{B}_1^{\star} = \mv{D}^{\star} + (1+\lambda^{\star})\mv{H},\label{B_1sum}\\
	\mv{B}_2^{\star} = \mv{D}^{\star}+(1+\gamma)\sum_{k=1}^K\beta_k^{\star}\mv{A}_k.\label{B_2sum}
\end{align}
We discuss the ranks of $\mv{W}^\star$ and $\mv{V}^\star$ in two cases. First, we introduce the following lemma.
\begin{lemma}\label{lemma_rank}
	Let $\mv{Y}$ and $\mv{X}$ be two matrices of the same dimension. It holds that ${\rm{rank}}(\mv{Y}+\mv{X})\geq {\rm{rank}}(\mv{Y})-{\rm{rank}}(\mv{X})$.
\end{lemma}
{\it{Proof:}} Please refer to \cite{bib4}.

\emph{Case I}: ${\rm{rank}}\left(\mv{D}^{\star}\right) = N_t$. Based on Lemma \ref{lemma_rank}, we have
\begin{align}
	{\rm{rank}}(\mv{B}_1^{\star})\geq {\rm{rank}}(\mv{D}^{\star})-1=N_t-1. 
\end{align}
If ${\rm{rank}}(\mv{B}_1^{\star}) = N_t$, we have $\mv{W}^{\star} = \mv{0}$, which cannot be the optimal solution to (P3.1R). Thus, we have ${\rm{rank}}(\mv{B}_1^{\star}) = N_t - 1$. We denote the null space of $\mv{B}_1^{\star}$ as $\mv{r}\in \mathbb{C}^{N_t\times1}$. Since ${\rm{rank}}(\mv{W}^{\star})=N_t - {\rm{rank}}(\mv{B}_1^{\star}) = 1$, it follows that $\mv{W}^{\star} = b\mv{r}\mv{r}^H,~b>0$, where $\mv{r}$ spans the null space of $\mv{B}_1^{\star}$.

\emph{Case II}: ${\rm{rank}}\left(\mv{D}^{\star}\right) = l <N_t$. In this case, we have $\mv{u}_n^H\mv{D}^{\star}\mv{u}_n = 0, n = 1,..., N_t - l$. According to (\ref{B_1sum}), we have 
\begin{align}
	\mv{u}_{n}^H\mv{B}_1^{\star}\mv{u}_n& = \mv{u}_{n}^H(\mv{D}^{\star} + (1+\lambda^{\star})\mv{H})\mv{u}_n\nonumber \\
	& =(1+\lambda^{\star})|\mv{u}_n^{H}\mv{h}|^2,~1\leq n\leq N_t - l.
\end{align} 
Since $\mv{B}_1^{\star}\preceq \mv{0}$ and $\lambda^{\star}>0$, $\mv{u}_n^{H}\mv{h}=0$ holds for $1\leq n\leq N_t - l$. According to (\ref{B_1sum}) and Lemma \ref{lemma_rank}, we have ${\rm{rank}}(\mv{B}_1^{\star}) \geq l -1$. We denote the null space of $\mv{B}_1^{\star}$ as $\mv{\Omega}$, which satisfies  ${\rm{rank}}(\mv{\Omega}) \leq N_t - l+1$. Since $\mv{u}_n^H\mv{B}_1^{\star}\mv{u}_n, n = 1,..., N_t-l$, we can get ${\rm{rank}}(\mv{B}_1^{\star})\geq N_t - l$. If ${\rm{rank}}(\mv{B}_1^{\star}) = N_t - l$, it follows that $\mv{\Omega} = \mv{U}$. Then, we can get $\mv{W}^{\star} = a_n\mv{u}_n\mv{u}_n^H, n = 1,..., N_t-l$ and $a_n\geq 0$. However, this implies that $\mv{u}_n^H\mv{h} = 0, n=1,..., N_t-1$, which cannot correspond to the optimal solution to (P3.1R). Thus, we have ${\rm{rank}}(\mv{\Omega})=N_t-l+1$, $\mv{\Omega}= [\mv{U}, \mv{r}]$, and ${\rm{rank}}(\mv{\Omega}) = N_t-l+1$. It then follows that 
\begin{align}
	\mv{W}^{\star}= b\mv{r}\mv{r}^H+ \sum_{n=1}^{N_t-l}a_n\mv{u}_n\mv{u}_n^{H},
\end{align}
where $b>0$, $a_n\geq 0$, and $\mv{r}$ is orthogonal to the span of $\mv{U}$, i.e., $\mv{r}^H\mv{U} = \mv{0}$. In this case, we prove that the solutions $(\tilde{\mv{W}}^{\star}, \tilde{\mv{V}}^{\star}, \tilde{t}^{\star})$ in (\ref{Ws}), (\ref{Vs}), and (\ref{ts}) constructed with a rank one solution of $\mv{W}$ is the optimal solution to (P3.1R). It can be shown that 
\begin{align}
	{\rm{tr}}\left(\mv{H}\tilde{\mv{W}}^{\star}\right) &= {\rm{tr}}\left(\mv{H}\left(\mv{W}^{\star}-\sum_{n=1}^{N_t-l}a_n\mv{u}_n\mv{u}_n^H\right)\right) \nonumber \\
	&= {\rm{tr}}\left(\mv{H}\mv{W}^{\star}\right),
\end{align}
\begin{align}
	{\rm{tr}}\left(\mv{H}\tilde{\mv{V}}^{\star}\right)+\tilde{t}^{\star}\sigma^2 &= {\rm{tr}}\left(\mv{H}\left(\mv{V}^{\star}+\sum_{n=1}^{N_t-l}a_n\mv{u}_n\mv{u}_n^H\right)\right)+t^{\star}\sigma^2\nonumber \\
	&= {\rm{tr}}\left(\mv{H}\mv{V}^{\star}\right) +t^{\star}\sigma^2=1,
\end{align}
\begin{align}
	&{\rm{tr}}\left({\mv{A}_k\tilde{\mv{W}}^{\star}}\right) = {\rm{tr}}\left(\left({\mv{A}_k\tilde{\mv{W}}^{\star}}-\sum_{k=1}^{N_t-l}a_n\mv{u}_n\mv{u}_n^H\right)\right)\nonumber\\
	&\leq {\rm{tr}}\left(\mv{A}_k\mv{W}^{\star}\right)
	\leq \gamma\left({\rm{tr}}\left(\mv{A}_k\mv{V}^{\star}\right)+\frac{\bar{t}^{\star}\sigma_{\mathrm{E}}^2r^2}{\bar{\beta}_0}\right)\nonumber\\
	&\leq \gamma\left({\rm{tr}}\left(\mv{A}_k\left(\mv{V}^{\star}+\sum_{n=1}^{N_t-l}a_n\mv{u}_n\mv{u}_n^H\right)\right)+\frac{\tilde{t}^{\star}\sigma_{\mathrm{E}}^2r^2}{\bar{\beta}_0}\right)\nonumber\\
	&=\gamma\left({\rm{tr}}\left(\mv{A}_k\tilde{\mv{V}}^{\star}\right)+\frac{t^{\star}\sigma_{\mathrm{E}}^2r^2}{\bar{\beta}_0}\right),
\end{align}
\begin{align}
	{\rm{tr}}\left(\tilde{\mv{W}}^{\star}+\tilde{\mv{V}}^{\star}\right)={\rm{tr}}\left({\mv{W}}^{\star}+{\mv{V}}^{\star}\right)\leq t^{\star}P,
\end{align}
\begin{align}
	& \begin{bmatrix}
		{\rm{tr}}\!\left(\mv{M}_2\!\left(\tilde{\mv{W}}^{\star}+\tilde{\mv{V}}^{\star}\right)\right)\!-\!\tilde{t}^{\star}\xi & {\rm{tr}}\left(\mv{M}_3\!\left(\tilde{\mv{W}}^{\star}+\tilde{\mv{V}}^{\star}\right)\right)\!\\
		{\rm{tr}}\!\left(\mv{M}_3^H\left(\tilde{\mv{W}}^{\star}+\tilde{\mv{V}}^{\star}\right)\!\right)\!& {\rm{tr}}\left(\mv{M}_1\!\left(\tilde{\mv{W}}^{\star}+\tilde{\mv{V}}^{\star}\right)\right)\!\\
	\end{bmatrix}\!\!=\!\nonumber\\
	&\begin{bmatrix}
		{\rm{tr}}\!\left(\mv{M}_2\!\left({\mv{W}}^{\star}+{\mv{V}}^{\star}\right)\right)\!-\!{t}^{\star}\xi & {\rm{tr}}\left(\mv{M}_3\!\left({\mv{W}}^{\star}+{\mv{V}}^{\star}\right)\right)\\
		{\rm{tr}}\!\left(\mv{M}_3^H\left({\mv{W}}^{\star}+{\mv{V}}^{\star}\right)\!\right)& {\rm{tr}}\left(\mv{M}_1\!\left({\mv{W}}^{\star}+{\mv{V}}^{\star}\right)\right)\\
	\end{bmatrix}\succeq \mv{0},
\end{align}
\begin{align}
	\tilde{\mv{W}}^{\star}\succeq \mv{0},\\
	\tilde{\mv{V}}^{\star}\succeq \mv{0},\\
	\tilde{t}^{\star}>0.
\end{align}
Hence, the constructed solution with a rank-one $\mv{W}$ is a feasible solution to (P3.1R) which achieves the same objective value. This thus completes the proof of Proposition \ref{prop_rank}.

\bibliographystyle{IEEEtran}
\bibliography{secure_ISAC}
\end{document}